# Theory of nonlinear terahertz susceptibility in ferroelectrics

Yujie Zhu[1], Taorui Chen[1], Aiden Ross[2], Bo Wang,[3] Xiangwei Guo[1], Venkatraman Gopalan[2], Long-Qing Chen[2], Jia-Mian Hu[1*]

[1]Department of Materials Science and Engineering, University of Wisconsin-Madison, Madison, WI, 53706, USA

[2]Department of Materials Science and Engineering, The Pennsylvania State University, University Park, Pennsylvania, 16802, USA

[3]Materials Science Division, Lawrence Livermore National Laboratory, Livermore, California 94550, USA

**Abstract**

An analytical theory is developed for predicting the nonlinear susceptibility of ionic polarization to continuous electromagnetic waves in both bulk and strained thin film ferroelectrics. Using a perturbation method for solving the nonlinear equation of motion for ionic polarization within the framework of Landau-Ginzburg-Devonshire theory, the full second-order nonlinear susceptibility tensor is derived as a function of frequency, temperature, and strain. The theory predicts the coexistence of a significantly enhanced second-order dielectric susceptibility and a relatively low dielectric loss in $BaTiO_3$ films with a strain-stabilized monoclinic ferroelectric phase and in a strained $SrTiO_3$ film near its temperature-driven second-order ferroelectric-to-paraelectric phase transition. This work establishes a theoretical framework for predicting and exploiting nonlinear interactions between THz waves and ferroelectric materials, and more generally, suggests exciting opportunities to strain-engineer nonlinear dynamical properties of ferroelectrics beyond the static and quasi-static limits.

*E-mail: jhu238@wisc.edu

## I. Introduction

Nonlinear susceptibility in the terahertz (THz) band (frequency: 0.1-10 THz) is critical to nonlinear THz wave interactions such as THz high-order harmonic generation [1–5], THz-field-induced second harmonic generation (SHG) [6–8], and THz field-induced Kerr effect [9]. These nonlinear processes underpin the development of a wide variety of THz applications ranging from nonlinear THz spectroscopy to high-power THz sources and to THz imaging [10,11].

Ferroelectric materials, due to their non-centrosymmetric nature, permit even-order harmonic generation and have been widely used for nonlinear wave phenomena at optical frequencies (~$10^{14}$ Hz). For example, $LiNbO_3$, a workhorse nonlinear optical material [12,13], is a uniaxial ferroelectric whose spontaneous polarization (denoted as $\mathbf{P}^{ion}$ to indicate its origin in ionic displacements) aligns along the $c$ axis of its hexagonal unit cell. More specifically, in displacive ferroelectric systems such as $LiNbO_3$, $PbTiO_3$ and $BaTiO_3$, the $\mathbf{P}^{ion}$ is caused by the condensation of a soft mode phonon below the Curie temperature and can be expressed as $\mathbf{P}^{ion}=Z_P^*\mathbf{Q}^P$ where $Z_P^*$ and $\mathbf{Q}^P$ are the Born effective charge density and the normal coordinates of the soft mode, respectively [14,15].

In the optical regime, frequency-dependent dynamic (linear or nonlinear) susceptibility of ferroelectric materials is primarily associated with the optical electric field-induced electronic polarization $\mathbf{P}^e$ [16,17] (note that $\mathbf{P}^e=0$ under zero electric field). In the THz and gigahertz (GHz) regimes, dynamic susceptibility of ferroelectrics contains contributions from both the $\mathbf{P}^{ion}$ and $\mathbf{P}^e$, because both types of polarizations can promptly respond to the GHz-THz electric fields given their high resonant frequencies (~$10^{12}$ Hz for $\mathbf{P}^{ion}$ [18] and ~$10^{15}$ Hz for $\mathbf{P}^e$). However, the contribution of $\mathbf{P}^{ion}$ to the dynamic susceptibility should be much more significant [19] because the linear dielectric susceptibility of $\mathbf{P}^{ion}$ is much larger than the induced $\mathbf{P}^e$ in the GHz-THz range, and because the $\mathbf{P}^{ion}$ can resonantly interact with the THz electric field.

Frequency-dependent nonlinear susceptibilities of $\mathbf{P}^{ion}$ in bulk ferroelectric crystals have been analytically calculated by employing the perturbation method to solve the equation of motion for $\mathbf{P}^{ion}$ within the framework of a Landau-Devonshire-type thermodynamic energy density function (a polynomial of $\mathbf{P}^{ion}$) [20–24]. However, the thermodynamic potential used in these works [20–24] does not incorporate the coupling between $\mathbf{P}^{ion}$ and strain (i.e., the piezoelectric effect), which has recently been shown to have a significant influence on the resonant frequency of the $\mathbf{P}^{ion}$ even in stress-free bulk ferroelectric materials [18,25]. Furthermore, these works [20–24] are largely focused on the mathematical derivation without extensively discussing the underlying physical picture, e.g., the relation between the nonlinear susceptibility and the landscape of the thermodynamic energy density.

In this article, based on the equation of motion for $\mathbf{P}^{ion}$ (specifically, the ionic polarization associated with the soft mode) and the Landau-Ginzburg-Devonshire (LGD) thermodynamical energy density function of $\mathbf{P}^{ion}$ that incorporates the coupling between $\mathbf{P}^{ion}$ and strain, we analytically derive the full nonlinear susceptibility tensor of $\mathbf{P}^{ion}$ in monodomain ferroelectrics and incipient ferroelectrics as a function of frequency, temperature, and, in the case of coherently strained thin films, the epitaxial strain.

While the theoretical framework presented in this paper is directly applicable to various types of THz nonlinear phenomena (e.g., third harmonic and sum/difference frequency generation) in ferroelectrics, the examples are focused on THz SHG, where a nonlinear THz polarization $\mathbf{P}^{ion}$ with an angular frequency $2\omega$ is generated in the material by an incident THz field $\mathbf{E}$ with an angular frequency $\omega$. After calculating the THz second-harmonic susceptibility in a few bulk ferroelectric single crystals, we predict a dramatically enhanced THz second-harmonic susceptibility in monodomain $BaTiO_3$ films with a strain-stabilized monoclinic phase, as well as near the temperature-driven ferroelectric-to-paraelectric phase transition in strained SrTiO$_3$ films. These findings indicate the potential application of using these strained ferroelectric films for source-current-free SHG at both the microwave and THz frequencies. The theory advances the physical understanding of the nonlinear interaction between THz waves and ferroelectrics. More broadly, this work suggests exciting opportunities for strain engineering of nonlinear dynamic properties in ferroelectrics beyond the static and quasi-static limits.

## II. Theory

The nonlinear THz susceptibilities refer to the third- and higher-rank $\chi$ tensors in the relation between the total ionic polarization $\mathbf{P}^{\text{ion}}$ (hereafter $\mathbf{P}$) and the incident electric field $\mathbf{E}$. In ferroelectric materials that have a spontaneous polarization $P_i^0$ under zero electric field, the total polarization can be written as,

$$P_i = P_i^0 + \Delta P_i = P_i^0 + \Delta P_i^{(1)} + \Delta P_i^{(2)} + \cdots, \tag{1a}$$

$$\Delta P_i^{(1)}(\omega) = \kappa_0 \sum_j \chi_{ij}^{(1)}(\omega) E_j(\omega), \tag{1b}$$

$$\Delta P_i^{(2)}(\omega_n \pm \omega_m) = \kappa_0 \sum_{jk} \sum_{(mn)} \chi_{ijk}^{(2)}(\omega_n \pm \omega_m, \omega_n, \omega_m) E_j(\omega_n) E_k(\omega_m), \tag{1c}$$

where $\kappa_0$ is the vacuum permittivity and $E_i$ is the electric-field component of the incident THz wave. The subscripts $i,j,k=1,2,3$ indicate crystal physics coordinates. Under the plane wave assumption in the thin slab limit, i.e., the thickness of the ferroelectric is significantly smaller than THz wavelength $\lambda_c = \frac{2\pi}{k^{\text{Re}}}$ ($k^{\text{Re}}$ is the real component of the complex wave number $\mathbf{k}$ [25]), one has $E_i = E_i^0 e^{-\mathbf{i}\omega t}$. Equation (1c) follows the notation in [26,27] except that all the $\mathbf{P}$ hereafter refers to ionic polarization; $\Delta P_i$ in Eq. (1a) is the dynamic variation of the ionic polarization, which can be separated into a first-order polarization $\Delta P_i^{(1)}(\omega) = \Delta P_i^{(1),0} e^{-\mathbf{i}\omega t}$ that has the same angular frequency as the incident THz wave (Eq. (1b)) and second-order polarization $\Delta P_i^{(2)}(\omega_n \pm \omega_m) = \Delta P_i^{(1),0} e^{-\mathbf{i}(\omega_n \pm \omega_m)t}$ that can contain multiple frequency components (Eq. (1c)). Here the bold "$\mathbf{i}$" denotes the imaginary unit. Specifically, an incident THz wave containing waves of two angular frequencies $\omega_n$ and $\omega_m$ (the notation $(mn)$ indicating that the $\omega_n \pm \omega_m$ is fixed while $\omega_n$ and $\omega_m$ can individually vary) can generate two SHG polarization components $\Delta P_i^{(2)}(2\omega_n)$ and $\Delta P_i^{(2)}(2\omega_m)$, a sum frequency generation (SFG) component $\Delta P_i^{(2)}(\omega_n + \omega_m)$, a difference frequency generation (DFG) component $\Delta P_i^{(2)}(\omega_n - \omega_m)$, and a dc polarization shift $\Delta P_i^{(2)}(0)$ induced by a static electric field rectified from complex electric fields $E_i(\omega) = E_i^0 e^{-\mathbf{i}\omega t}$ and its conjugate $E_i^*(\omega) = E_i(-\omega) = E_i^0 e^{\mathbf{i}\omega t}$.

Each second-order polarization component is associated with its own $\chi_{ijk}^{(2)}$ tensor, including $\chi_{ijk}^{(2)}(2\omega_n, \omega_n, \omega_n)$, $\chi_{ijk}^{(2)}(2\omega_m, \omega_m, \omega_m)$, $\chi_{ijk}^{(2)}(\omega_n + \omega_m, \omega_n, \omega_m)$, $\chi_{ijk}^{(2)}(\omega_n - \omega_m, \omega_n, -\omega_m)$, and $\chi_{ijk}^{(2)}(0, \omega_m, -\omega_m)$ and/or $\chi_{ijk}^{(2)}(0, \omega_n, -\omega_n)$. In this work, the analytical formulae for all these $\chi_{ijk}^{(2)}$ tensors are derived. The examples focus on the THz SHG, where a monochromatic incident THz wave with an angular frequency $\omega$, $\mathbf{E}(\omega)$, generates a second-order polarization $\Delta P_i^{(2)}(2\omega)$, i.e.,

$$\Delta P_i^{(2)}(2\omega) = \kappa_0 \sum_{jk} \chi_{ijk}^{(2)}(2\omega, \omega, \omega) E_j(\omega) E_k(\omega). \tag{2}$$

The analytical formulae for the frequency-dependent linear susceptibility $\chi_{ij}^{(1)}$ and second-order susceptibility $\chi_{ijk}^{(2)}$ can be obtained by finding the steady-state solution of the equation of motion for the $\Delta P_i$ [25], which is analogous to the equation of motion for an anharmonic oscillator, given as,

$$\mu \frac{\partial^2 \Delta P_i}{\partial t^2} + \gamma_{ij} \frac{\partial \Delta P_j}{\partial t} = \Delta E_i^{\text{eff}} = E_i^{\text{Landau}} + E_i^{\text{Elast}} + E_i^{\text{d}} + E_i + E_i^{\text{rad}}, \quad (3)$$

where the subscripts $i$=1,2,3 of the polarization component indicate the Cartesian crystal physics coordinates of the paraelectric cubic phase within the framework of the LGD theory [28]. $\mu = \frac{1}{\kappa_0 \omega_p^2}$ is the mass coefficient (polarization inertia) with an ionic plasma frequency $\omega_p = \sqrt{\frac{1}{\kappa_0 V_0} \sum_n \frac{q_n^2}{M_n}}$, where $q_n$ and $M_n$ are the charge and mass of the $n$th charged ion in a unit cell with a volume $V_0$ [29], and $\gamma_{ij}$ is the phenomenological viscous damping coefficient that can be related to the crystal viscosity [30]; the temporal variation of the total effective electric field is $\Delta E_i^{\text{eff}} = E_i^{\text{eff}}(P_i) - E_i^{\text{eff}}(P_i^0)$. At the initial equilibrium state ($P_i = P_i^0$), one has $E_i^{\text{eff}}(P_i^0) = 0$ and thus, $\Delta E_i^{\text{eff}} = E_i^{\text{eff}}(P_i)$. Among the various (effective) electric fields contributing to the $E_i^{\text{eff}}$, $E_i^{\text{Landau}} = -\frac{\partial f^{\text{Landau}}}{\partial P_i}$ and $E_i^{\text{Elast}} = -\frac{\partial f^{\text{Elast}}}{\partial P_i}$ are nonlinear polynomials of $P_i$ ($f^{\text{Landau}}$ and $f^{\text{Elast}}$ are the Landau and elastic energy densities, respectively) [25]. The derivation of Eq. (3) is provided in Appendix A.

Specific ferroelectric materials considered in this work include (i) tetragonal perovskite ferroelectric single crystals BaTiO$_3$ and PbTiO$_3$ where the initial equilibrium polarization **P**$^0$ aligns along the $x$ axis in the lab coordinate system; (ii) trigonal ferroelectric single crystals LiTaO$_3$ and LiNbO$_3$ where the initial equilibrium polarization **P**$^0$ also aligns along the $x$ axis in the lab coordinate system; (iii) an anisotropically strained (001)$_{\text{pc}}$ BaTiO$_3$ (pc: pseudocubic) thin film which has a surface parallel to the (001) plane of its cubic paraelectric phase; and (iv) a biaxially strained (001)$_{\text{pc}}$ SrTiO$_3$ film grown on an orthorhombic (o) DyScO$_3$ (110)$_{\text{o}}$ substrate [31]. Normal incidence of a $p$-polarized THz wave is considered under the plane wave assumption, where the THz electric field inside the ferroelectric material, the $E_i$ in Eq. (3), only contains an $x$ component in the lab coordinate system. The lab and the crystal physics coordinate systems in the above four cases are illustrated in Fig. 1. Expressions of $f^{\text{Landau}}$ and $f^{\text{Elast}}$ for BaTiO$_3$, PbTiO$_3$, SrTiO$_3$, LiTaO$_3$ and LiNbO$_3$ and the relevant materials parameters are summarized in Appendix B.

Under the plane wave assumption in the thin slab limit (i.e., the thickness of the ferroelectric slab is much smaller than the THz wavelength in the slab) [23], the polarization-oscillation-induced radiation electric field can be calculated as $E_i^{\text{rad}} = (-\frac{d_0}{2\kappa_0 c} \frac{\partial P_x}{\partial t}, -\frac{d_0}{2\kappa_0 c} \frac{\partial P_y}{\partial t}, 0)$ for a single-domain ferroelectric [25], where $d_0$ is the thickness of the ferroelectric and $c$ is the speed of light in vacuum. For thick bulk crystals, $E_i^{\text{rad}}$ has a complex analytical expression and varies spatially along the thickness direction, as derived in [25], At the initial equilibrium state, both the $E_i$ and $E_i^{\text{rad}}$ are zero. For a single-domain ferroelectric thin film with an infinitely large $x$-$y$ plane and mobile screening charges (e.g., electrons, holes, and absorbed ions) at the top and bottom surfaces, the depolarization field at the initial equilibrium state is zero. Under the excitation by THz or higher-frequency electric fields, we assume that these charged species at the surfaces remain largely frozen. Consequently, the film is subjected to a dynamic depolarization field that is given by $\Delta E_i^{\text{d}} = (0, 0, -\frac{1}{\kappa_0 \kappa_b} \Delta P_z)$, where $\kappa_b$ is the background dielectric constant accounting for the contribution from the electronic contribution [32,33]. Here, a typical value of $\kappa_b$=5 [34] is used for all ferroelectric materials in the calculation. The lab coordinate system ($i=x,y,z$) is used in the above expressions for the $E_i$, $E_i^{\text{rad}}$, and $E_i^{\text{d}}$.

The elastic energy $f^{\text{Elast}} = \frac{1}{2}c_{ijkl}(\varepsilon_{kl} - \varepsilon_{kl}^0)(\varepsilon_{ij} - \varepsilon_{ij}^0)$, where $c_{ijkl}$ is the elastic stiffness tensor under constant electric field and temperature; the stress-free strain $\boldsymbol{\varepsilon}^0$ is induced by the $P_i$ through the electrostrictive effect. The solution of the total strain $\varepsilon_{ij}$ at the initial equilibrium state depends on the mechanical boundary condition of the system. In the case of coherently strained BaTiO$_3$ and SrTO$_3$ thin films, one has a mixed boundary condition [35] with $\varepsilon_{11} = \varepsilon_{11}^{\text{mis}}$, $\varepsilon_{22} = \varepsilon_{22}^{\text{mis}}$, $\varepsilon_{12} = 0$ and $\sigma_{i3} = 0$ ($i$=1,2,3) from which the expressions of the total strain $\varepsilon_{i3}$ can be obtained. The mismatch strain $\varepsilon_{11}^{\text{mis}}$ and $\varepsilon_{22}^{\text{mis}}$ result from the lattice constant and/or thermal expansion coefficient mismatch between the epitaxial film and substrate. It is known that such mismatch strain can enable polymorphic ferroelectric phase transitions that are absent in the stress-free state or modulate the ferroelectric-to-paraelectric transition temperature in ferroelectric thin films [35,36] and that the static (i.e., driven by dc electric field) linear dielectric susceptibility $\chi_{ij}^{(1),\text{dc}} = \frac{\partial P_i}{\kappa_0 \partial E_j}$ can be significantly enhanced near such phase transitions [28,37].

To analytically solve the $\chi_{ijk}^{(2)}$, we first rewrite Eq. (3) into the following matrix form by expanding the $E_i^{\text{Landau}}$ and $E_i^{\text{Elast}}$ in Taylor series and dropping the higher-order terms (see Appendix C),

$$\mu \frac{\partial^2 \Delta \mathbf{P}}{\partial t^2} + \boldsymbol{\gamma}^{\text{eff}} \frac{\partial \Delta \mathbf{P}}{\partial t} + \mathbf{K}\Delta \mathbf{P} + \mathbf{C}\Delta \mathbf{P}_{\text{II}} = \mathbf{E}, \qquad (4)$$

where both the polarization change $\Delta \mathbf{P} = (\Delta P_1, \Delta P_2, \Delta P_3)^{\text{T}}$ and the incident THz electric field $\mathbf{E} = (E_1, E_2, E_3)^{\text{T}}$ are a 3×1 matrix; $\Delta \mathbf{P}_{\text{II}} = (\Delta P_1^2, \Delta P_2^2, \Delta P_3^2, 2\Delta P_1 \Delta P_2, 2\Delta P_1 \Delta P_3, 2\Delta P_1 \Delta P_2)^{\text{T}}$ is a 6×1 matrix; $\boldsymbol{\gamma}^{\text{eff}}$ is a 3×3 matrix which contains contributions from both the phenomenological intrinsic damping (related to crystal viscosity of the ferroelectric) and the radiation-induced damping [18,25]. For strained ferroelectric thin films in the thin slab limit, one has

$$\boldsymbol{\gamma}^{\text{eff}} \approx \begin{bmatrix} \gamma_{11} + \frac{1}{2}\frac{d_0}{\kappa_0 c} & 0 & 0 \\ 0 & \gamma_{22} + \frac{1}{2}\frac{d_0}{\kappa_0 c} & 0 \\ 0 & 0 & \gamma_{33} \end{bmatrix}, \qquad (5)$$

where the off-diagonal components are assumed to be zero and the intrinsic damping coefficients are assumed to be isotropic $\gamma_{11} = \gamma_{22} = \gamma_{33} = 2 \times 10^{-7}$ Ω·m [25]. The additional term added to $\gamma_{11}$ and $\gamma_{22}$ describes the radiation-electric-field-induced damping [18], where the thickness of the ferroelectric film $d_0$=10 nm is small enough to ensure that the film is coherently strained by the substrate. For thick bulk crystals, the complex expression of the radiation electric field $E_i^{\text{rad}}$ would result in a radiation-induced damping that varies along the thickness and does not have an explicit analytical expression [23]. For simplicity, we use the same $\boldsymbol{\gamma}^{\text{eff}}$ in the analytical calculation of both bulk and thin-film ferroelectrics in this work. For the components of the 3×3 matrix $\mathbf{K}$, $K_{ij} = -A_{ij} - B_{ij}$ ($i,j$=1,2,3). In the case of (001)$_{\text{pc}}$ BaTiO$_3$ and SrTiO$_3$ thin films (Figs. 1(c-d)) under dynamic electric-field excitation ($\omega \neq 0$), we use $K_{33} = -A_{33} - B_{33} + \frac{1}{\kappa_0 \kappa_{\text{b}}}$, where the term $1/\kappa_0 \kappa_{\text{b}}$ results from the dynamic depolarization field $\Delta E_z^{\text{d}}(t) = -\Delta P_z(t)/\kappa_0 \kappa_{\text{b}}$ ($z \equiv 3$) in a thin film with infinitely large $x$-$y$ plane (see Appendix B). In the case of dc excitation ($\omega$=0) and/or bulk ferroelectric crystals, the dynamic depolarization field does not need to be considered, thus $K_{33}=-A_{33}-B_{33}$. For the components of the third-rank tensor **C**, $C_{ijk} = -\frac{1}{2}A_{ijk} - \frac{1}{2}B_{ijk}$ ($i,j,k$=1,2,3), where,

$$A_{ij} = -\frac{\partial^2 f^{\text{Landau}}}{\partial P_i \partial P_j}\bigg|_{\mathbf{P}=\mathbf{P}^0}, A_{ijk} = -\frac{\partial^3 f^{\text{Landau}}}{\partial P_i \partial P_j \partial P_k}\bigg|_{\mathbf{P}=\mathbf{P}^0}, \quad (6a)$$

$$B_{ij} = -\frac{\partial^2 f^{\text{Elas}}}{\partial P_i \partial P_j}\bigg|_{\mathbf{P}=\mathbf{P}^0}, B_{ijk} = -\frac{\partial^3 f^{\text{Elas}}}{\partial P_i \partial P_j \partial P_k}\bigg|_{\mathbf{P}=\mathbf{P}^0}. \quad (6b)$$

Thus, $K_{ij}$ and $C_{ijk}$ represent, respectively, the local curvature of the total free energy density (a sum of the Landau and elastic energy density) at the initial equilibrium (spontaneous) polarization state $\mathbf{P}^0$. Both the mismatch strain and the temperature can modulate the $\mathbf{P}^0$, **K**, and **C** tensors.

We then employ the perturbation method, which has previously been used to derive the nonlinear susceptibility of electronic polarization at optical frequencies [26], to analytically solve Eq. (3) for both the linear and nonlinear susceptibility (see Appendix C). The linear susceptibility $\chi_{ij}^{(1)}(\omega)$ is given as,

$$\chi_{ij}^{(1)}(\omega) = \frac{1}{\kappa_0}\begin{bmatrix} D_{11}(\omega) & K_{12} & K_{13} \\ K_{21} & D_{22}(\omega) & K_{23} \\ K_{31} & K_{32} & D_{33}(\omega) \end{bmatrix}^{-1}, i,j = 1,2,3, \quad (7)$$

In Eq. (4), the diagonal components $D_{ii}(\omega) = \mu(\omega_i^2 - \omega^2) - \mathbf{i}\gamma_{ii}^{\text{eff}}\omega$, where the resonant frequency of polarization oscillation $\omega_i = \sqrt{K_{ii}/\mu}$ [25]. Under the application of a static (dc) electric field ($\omega$=0), one has $\chi_{ij}^{(1),\text{dc}} = 1/\kappa_0 K_{ij}$ ($i,j$=1,2,3).

The THz SHG susceptibility $\chi_{ijk}^{(2)}(2\omega,\omega,\omega)$ is given by,

$$\chi_{ijk}^{(2)}(2\omega,\omega,\omega) = -\kappa_0^2 \sum_{\alpha,\beta,\gamma=1,2,3} C_{\alpha\beta\gamma}\chi_{i\alpha}^{(1)}(2\omega)\chi_{\beta j}^{(1)}(\omega)\chi_{\gamma k}^{(1)}(\omega). i,j,k = 1,2,3 \quad (8)$$

Equation (8) indicates that $\chi_{ijk}^{(2)}$ can be expressed as a function of linear susceptibilities $\chi_{ij}^{(1)}$, which is consistent with the theory by Garret [38] and later by Mayer and Keilmann [39]. A notable finding of our theory is that the coefficient $C_{\alpha\beta\gamma}$, which was referred to as generalized Miller's coefficient [39] and typically fitted to experimental measurement [39,40], is now specifically connected to the third-order derivatives of the LGD energy density with respect to the equilibrium polarization $\mathbf{P}^0$. For SHG, the tensor $d_{ijk} = \frac{1}{2}\chi_{ijk}^{(2)}$ is also used. In this paper, we use $\chi_{ijk}^{(2)}$ to show its relation with $\chi_{ij}^{(1)}$, as in Eq. (8).

To demonstrate the validity of Eq. (8), two tests are performed (see details in Appendix D). First, the expression of $\chi_{ijk}^{(2),\text{dc}}$ directly from thermodynamic analysis is the same as the expression obtained by letting $\omega$=0 in Eq. (7). As an example, in the case of a tetragonal BaTiO$_3$ bulk crystal, we first calculate the static $\chi_{333}^{(2),\text{dc}}$ based on Eq. (7), which describes the generation of static nonlinear polarization $\Delta P_3$ by a static electric field $E_3$ ($\equiv E_x$ in the lab coordinate system, see Fig. 1(a)). The calculate value ($\chi_{333}^{(2),\text{dc}}$=-1.573×10$^{-6}$ m/V) agrees well with the value of -1.576×10$^{-6}$ m/V extracted by fitting a static $P_3$–$E_3$ curve obtained from thermodynamic analysis. Second, the numbers of nonzero and independent elements in the third-rank tensor $\chi_{ijk}^{(2),\text{dc}}$ for stress-free BaTiO$_3$ crystals of cubic, tetragonal, orthorhombic, or rhombohedral phase are consistent with those of the second-order susceptibility tensor of the electronic polarization (optical SHG) under the same crystal symmetry [26].

**III. Results and Discussion**

We first calculate the frequency-dependent nonlinear susceptibility at room temperature (25°C) in the bulk tetragonal BaTiO$_3$ and PbTiO$_3$ as well as the trigonal LiNbO$_3$ and LiTaO$_3$ single crystals. As shown in Figs. 1(a) and 1(b), the initial equilibrium polarization $\mathbf{P}^0$ aligns along the polar axis of the BaTiO$_3$, PbTiO$_3$, LiNbO$_3$ and LiTaO$_3$ in the crystal physics coordinates, which is the $x$ axis in the lab coordinate system. When the incident THz wave is polarized only along the $x$ in the lab coordinate system ($E_x \equiv E_3$), one has,

$$\Delta P_3^{(2)}(2\omega) = \kappa_0 \chi_{333}^{(2)} E_3(\omega)^2, \tag{9}$$

Since the amplitude of the incident THz electric field inside the material has a time-dependence, $E_3(\omega) = E_3^0 e^{-i\omega t}$, the amplitude and the phase of the second-order nonlinear polarization, $\Delta P_3^{(2)}(2\omega) = \Delta P_3^{(2),0} e^{i(-2\omega t + \varphi^{(2)})}$ are related to the modulus, $\left|\chi_{iii}^{(2)}\right|$, and the argument, $\theta$, of the $\chi_{ijk}^{(2)}$, respectively, i.e.,

$$\Delta P_3^{(2),0} = \frac{\kappa_0}{2}\left|\chi_{333}^{(2)}\right|E_3^{0^2}, \quad \varphi^{(2)} = \theta, \tag{10}$$

Figure 2(a) shows the frequency-dependent $\left|\chi_{333}^{(2)}\right|$ for the four ferroelectric materials. Notably, in the case of LiTaO$_3$, by tuning the Landau parameters provided in [41], good agreement with the experimental measurement [39] is achieved both in the nonlinear susceptibility modules $\left|\chi_{333}^{(2)}\right|$ and the resonant frequencies. Based on Eq. (7), one has,

$$\chi_{333}^{(2)} = -\kappa_0^2 C_{333} \chi_{33}^{(1)}(2\omega)\chi_{33}^{(1)}(\omega)^2. \tag{11}$$

Equation (11) also suggests the existence of two peaks for the $\left|\chi_{333}^{(2)}\right|$ at $\omega = \omega_3, \omega_3/2$, where the $\chi_{33}^{(1)}(\omega)$ and $\chi_{33}^{(1)}(2\omega)$ reach their maximum, respectively, as shown in Fig. 2(a). By comparison, the $\chi_{33}^{(1)}(\omega)$ resonates only at $\omega_3$, as shown in Fig. 2(b). In particular, the $\left|\chi_{333}^{(2)}\right|$ at $\omega_3/2$ is at about the same order-of-magnitude with its value at $\omega_3$. Furthermore, the dielectric loss of the ferroelectric [42], which is represented by the imaginary part of the linear susceptibility $\chi_{33}^{(1),\text{Im}}$, is three orders-of-magnitude smaller at $\omega_3/2$ (see bottom panel of Fig. 2(b)). Therefore, for potential applications of THz SHG, it is an attractive option to set the frequency of the incident THz wave at the half-resonance frequency of the ionic polarization in ferroelectrics.

By letting $\omega=0$ in Eq. (11), one has $\chi_{333}^{(2),\text{dc}} = -\kappa_0^2 C_{333} \chi_{33}^{(1),\text{dc}^3}$. Therefore, materials with large dc dielectric susceptibility $\chi_{33}^{(1),\text{dc}}$ also tend to have large $\chi_{333}^{(2),\text{dc}}$. From Fig. 2(a), it is evident that the tetragonal BaTiO$_3$ has a substantially larger $\chi_{333}^{(2),\text{dc}}$ and larger peak values of $\left|\chi_{333}^{(2)}\right|$ than the other three ferroelectric materials. As shown in Figs. 2(c), the local curvature of the energy landscape near the $P_3^0$ is the smallest in the tetragonal BaTiO$_3$. As a result, the tetragonal BaTiO$_3$ has the largest $\chi_{33}^{(1),\text{dc}}$ and hence the $\left|\chi_{333}^{(2),\text{dc}}\right|$ among the four materials, as shown in Fig. 2(d). Furthermore, the analytically calculated $\left|\chi_{333}^{(2)}\right|$ of the tetragonal BaTiO$_3$ agrees well with the values extracted independently from dynamical phase-field simulations (see Appendix E), demonstrating that the analytical model is valid.

We now calculate the $\chi_{111}^{(2)}$ in an anisotropically strained $(001)_{pc}$ BaTiO$_3$ film at room temperature (25°C). Here, the $\chi_{111}^{(2)}$ is associated with the generation of second-order nonlinear polarization $\Delta P_1$ by a dynamic electric field $E_1$ ($\equiv E_x$ in the lab coordinate system, see Fig. 1(c)). The BaTiO$_3$ film is subjected to a fixed mismatch strain $\varepsilon_{22}^{mis}$ = -1% yet the strain $\varepsilon_{11}^{mis}$ can vary. This strain condition is considered for three reasons. First, varying the $\varepsilon_{11}^{mis}$ from 2% to -1% leads to a transition from an in-plane tetragonal $T_1$ phase with ($P_1^0 \neq 0$, 0, 0) to an out-of-plane orthorhombic $O_{13}$ phase with ($P_1^0 \neq 0$, 0, $P_3^0 \neq 0$), followed by a transition to an out-of-plane tetragonal $T_3$ phase with (0, 0, $P_3^0 \neq 0$), as shown in Fig. 3(a). We can therefore study how these two typical polymorphic ferroelectric phase transitions influence the $\chi_{111}^{(2)}$. Second, the zero $P_2^0$ component in such an anisotropically strained film allows for excluding the contribution of $\chi_{2i}^{(1)}$ ($i$=1,2,3) to the $\chi_{111}^{(2)}$ (see Eq. (7)), thereby simplifying the analysis. Third, the three strain-stabilized polar phases ($T_1$, $O_{13}$, and $T_3$) have all been experimentally observed in BaTiO$_3$, where the $O_{13}$ phase is also defined as a monoclinic $M_C$ phase if $|P_3^0| > |P_1^0|$ [43,44].

The variation of $\chi_{111}^{(2),dc}$ with the $\varepsilon_{11}^{mis}$, as shown in Fig. 3(b), can be understood by analyzing the strain modulation of the local curvature and asymmetry of the energy landscape. For the tetragonal $T_1$ and $T_3$ phases, one can analogously derive that $\chi_{111}^{(2),dc} = -\kappa_0^2 C_{111} \chi_{11}^{(1),dc^3}$. In the $T_3$ phase, $C_{111} = 0$ since the local energy landscape is symmetric with respect to $\Delta P_1$, thus $\chi_{111}^{(2),dc}$=0. In the $T_1$ phase, the $\chi_{111}^{(2),dc}$ decreases as the $\varepsilon_{11}^{mis}$ increases, which is attributed to the decreasing $\chi_{11}^{(1),dc}$, as shown in Fig. 3(c). For the $O_{13}$ phase, a

$$\chi_{111}^{(2),dc} = -\kappa_0^2 \left( C_{111} \chi_{11}^{(1),dc^3} + 3C_{113} \chi_{11}^{(1),dc^2} \chi_{13}^{(1),dc} + 3C_{133} \chi_{11}^{(1),dc} \chi_{13}^{(1),dc^2} + C_{333} \chi_{13}^{(1),dc^3} \right). \quad (12)$$

As shown in Fig. 3(c), the diagonal component $\chi_{11}^{(1),dc}$ is much larger than the $\chi_{13}^{(1),dc}$ in the $M_C$ phase, especially near the $O_{13}$/$T_3$ phase boundary. Thus, the significant increase in $\chi_{111}^{(2),dc}$ of the $M_C$ phase is mainly caused by the associated increase in the $\chi_{11}^{(1),dc}$.

Let us now discuss the frequency dependence of the $|\chi_{111}^{(2)}|$ for the $O_{13}$ and $T_1$ phases under different strain $\varepsilon_{xx}^{mis}$, noting that $|\chi_{111}^{(2)}|$=0 in the $T_1$ phase. The $\chi_{111}^{(2)}$ of the tetragonal $T_1$ phase is given by,

$$\chi_{111}^{(2)} = -\kappa_0^2 C_{111} \chi_{11}^{(1)}(2\omega) \chi_{11}^{(1)}(\omega)^2. \quad (13)$$

Notably, Equation (13) above is also approximately applicable to the $O_{13}$ phase, since it is reasonable to consider $\chi_{i3}^{(1)} = \chi_{3i}^{(1)} \approx 0$ because the large dynamical depolarization field tends to suppress the magnitude of the out-of-plane polarization variation (i.e., $|\Delta P_3(t)|$ is much smaller than $|\Delta P_1(t)|$). Based on Eq. (13), there should be two peaks at $\omega_1$, $\omega_1/2$, in the frequency spectrum of $|\chi_{xxx}^{(2)}|$, consistent with the results in Fig. 3(d). The locations of the $\omega_1$ can be seen more clearly at the peaks in the frequency spectrum of $\chi_{11}^{(1),Im}$, as shown in Fig. 3(e).

As the strain $\varepsilon_{11}^{mis}$ becomes less compressive (e.g., from -0.05% to -0.008%) in the $O_{13}$ phase, $K_{11}$ decreases (see Fig. 3(c)), leading to a smaller $\omega_1 = \sqrt{K_{11}/\mu}$. As a result, the discrepancy between $\omega_1$ and $\omega_1/2$ also decreases. Since the peaks of $\omega_1$, $\omega_1/2$ both have a finite linewidth due to the nonzero damping, these two peaks can partially overlap when they are close. This explain why the $|\chi_{111}^{(2)}|$ at $\omega_1/2$ is larger than its value at $\omega_1/2$ at $\varepsilon_{11}^{mis}$=-0.008%. Together, Figure 3(d-e) demonstrate the effectiveness of using strain to enhance the $\left|\chi_{111}^{(2)}\right|$ yet to keep the $\chi_{11}^{(1),\text{Im}}$ (dielectric loss) at a relatively low value. Specifically, at $\varepsilon_{11}^{\text{mis}}=-0.008\%$, $\left|\chi_{111}^{(2)}\right|$ reaches a value of $\sim 2.54\times10^{-3}$ m/V at 0.45 THz, yet $\chi_{xx}^{(1),\text{Im}}$ is $\sim 636.5$ (one order of magnitude smaller than its peak value). This value of $\left|\chi_{111}^{(2)}\right|$ is one order of magnitude larger than the $\left|\chi_{\text{eff}}^{(2)}\right|$ ($\sim 10^{-4}$ m/V) reported in the superconducting NbN thin film [45].

The above analyses indicate that a significant enhancement in $\chi_{111}^{(2)}$ simultaneously requires a vanishing curvature (in other words, a large dielectric susceptibility) and a non-vanishing asymmetry ($C_{111}\neq 0$) of the energy landscape, which can be achieved in BaTiO$_3$ with a monoclinic $M_C$ phase, occurring near the strain-driven second-order $O_{13}-T_3$ ferroelectric phase transition in coherently strained BaTiO$_3$ film. With this understanding, we further evaluate the $\chi_{111}^{(2)}$ near the second-order ferroelectric-to-paraelectric transition in an equixially strained (001)$_{\text{pc}}$ SrTiO$_3$ film with $\varepsilon_{11}^{\text{mis}}=\varepsilon_{22}^{\text{mis}}=1\%$. As shown in Fig. 4(a), the calculated equilibrium polarization $\mathbf{P}^0$ is ($P_1^0$,0,0), which is consistent with the recent experimental observation in a coherently SrTiO$_3$ film grown on (110)$_\text{o}$ DyScO$_3$ substrate with a largely single polarization domain after in-plane electric poling [31]. As the temperature $T$ approaches the Curie point $T_\text{c}$, $P_1^0$ gradually decreases to zero (Fig. 4(a)); the $\chi_{111}^{(2),\text{dc}}=-\kappa_0^2 C_{111}\chi_{11}^{(1),\text{dc}^3}$ increases dramatically and then drops to zero in the paraelectric phase, as shown in Fig. 4(b). Likewise, this is mainly because $\chi_{11}^{(1),\text{dc}}$ is enhanced significantly at near the $T_\text{c}$ (see Fig. 4(c)) and because $C_{111}$ is zero in the cubic paraelectric phase.

The frequency spectrum of the $\left|\chi_{111}^{(2)}\right|$ of the in-plane tetragonal SrTiO$_3$ film should likewise only display two peaks at $\omega_1/2$ and $\omega_1$, which are tens of GHz due to the reduced soft mode frequency near the $T_\text{c}$ [46]. Furthermore, at close to $T_\text{c}$, the two peaks of the $\left|\chi_{xxx}^{(2)}\right|$ spectrum at $\omega_1/2$ and $\omega_1$ can merge into one, see for example the cases of 243.7 K and 243.5 K in Fig. 4(d). Importantly, these features allow for identifying the frequency of the incident THz/GHz wave for obtaining significantly enhanced $\left|\chi_{111}^{(2)}\right|$ together with a relatively low $\chi_{11}^{(1),\text{Im}}$. For example, at $T$=242 K, $\left|\chi_{111}^{(2)}\right|$ has a value of 0.466 m/V yet the $\chi_{11}^{(1),\text{Im}}$ is about 56.46 at 1 GHz; at $T$=242 K, $\left|\chi_{111}^{(2)}\right|$ has a value of 0.611 m/V at 16 GHz yet the $\chi_{11}^{(1),\text{Im}}$ is about 985. These values of $\left|\chi_{111}^{(2)}\right|$ are three orders of magnitude larger than the $\left|\chi_{\text{eff}}^{(2)}\right|$ ($\sim 10^{-4}$ m/V) reported in the superconducting NbN thin film [45]. Such a high $\left|\chi^{(2)}\right|$, along with manageable dielectric loss, suggests an exciting prospect of using coherently strained SrTiO$_3$ film as a structurally simple, source-current-free (and hence ultralow power dissipation) frequency doubler operating in the GHz/millimeter-wave band for high-data-rate wireless communication.

## IV. Conclusions

We have developed an analytical theory for predicting the dynamic nonlinear dielectric susceptibility of monodomain ferroelectric crystals as a function of frequency, temperature, and in the case of strained thin films, the epitaxial strain. Our theory reveals the important role of the strain-polarization coupling in ferroelectrics, which has been ignored in existing theoretical works [20–24], in determining the nonlinear dielectric susceptibility through the modulation of the curvature and asymmetry of the local energy landscape.

Based on the well-established LGD thermodynamic energy density function and the kinetic parameters of different ferroelectric materials, the theory predicts a route to enhancing the modulus of the second-harmonic susceptibility $\chi_{111}^{(2)}$ and simultaneously maintaining the dielectric loss at a low level in a (001)$_{\text{pc}}$

BaTiO$_3$ film with strain-stabilized monoclinic $M_C$ phase and a strained (001)$_{pc}$ SrTiO$_3$ film near its temperature-driven second-order ferroelectric-to-paraelectric phase transition. These results reveal the critical importance of stabilizing the $M_C$ phase in enhancing the $\chi_{111}^{(2)}$ of BaTiO$_3$ and similar ferroelectric systems, which is analogous to the critical role of $M_C$ phase in enhancing the nonlinear optical and piezoelectric property coefficients of BaTiO$_3$ [43,44]. In addition to the $\chi_{111}^{(2)}$ which is relevant to the THz SHG, the analytical formulae of other second-order $\chi_{ijk}^{(2)}$, including THz SFG/DFG and THz wave rectification (dc shift), are also derived (see Appendix D). By comparing the predicted $\chi_{ijk}^{(2)}$ to experimental measurements (e.g., THz SHG), one can refine the LGD coefficients (e.g., here we refine the coefficients of LiTaO$_3$, as shown in Fig. 1(a)) and the kinetic parameters such as the mass and damping coefficients of a wide range of ferroelectric materials. The procedures of derivations can be readily extended to calculate the higher-order susceptibilities (e.g., the $\chi_{ijkl}^{(3)}$) as well.

Overall, this work provides a theoretical basis for studying the nonlinear interaction between a THz (typical frequency range: 0.1-10 THz) or lower-frequency electromagnetic wave and a ferroelectric material. The analytically calculated nonlinear susceptibility of the ferroelectric polarization **P**$^{ion}$ (soft mode) is obtained by finding the *steady-state* solution to the equation of motion for **P**$^{ion}$, and thus only applicable to continuous THz wave or multi-cycle narrowband THz pulse. However, the theoretical framework can be readily generalized to calculate the nonlinear susceptibility of ferroelectric polarization to single-cycle broadband THz pulse, as used in many experiments [47–51], by finding the *transient-state* solution to the equation of motion for **P**$^{ion}$.

Moreover, by (i) extending the LGD potential (a polynomial of **P**$^{ion}$) to incorporate the coupling between **P**$^{ion}$ and the ionic polarization associated with the higher-frequency infrared (IR) active phonons (denoted as **P**$^{ion,IR}$), and (ii) solving the coupled equations of motion for both the **P**$^{ion}$ and **P**$^{ion,IR}$, it is possible to analytically calculate the nonlinear susceptibility of **P**$^{ion}$ to a mid-IR (frequency: ~20 THz) pulse, as has been studied in many computational [52–54] and experimental [14,55–57] studies (a.k.a. the nonlinear phononics approach). Furthermore, by introducing a nonlinear coupling between **P**$^{ion}$ and the electric field of a near-IR pulse (frequency: tens of THz) in the framework of impulsive stimulated Raman Scattering (ISRS) into the equation of motion for **P**$^{ion}$ (similarly to works [58–61]), it is possible to analytically calculate the nonlinear susceptibility of **P**$^{ion}$ to a near IR pulse via the ISRS.

More broadly, the theory in this work can be the susceptibilities under the excitation of a circularly polarized THz wave which can enable emergent phenomena such as the dynamic multiferroicity [15,62,63], to ferroelectrics under mechanical boundary conditions such as an uniaxially stretched ferroelectric membrane [64], and to other polar materials that have a spontaneous ionic polarization such as wurtzite III-nitride semiconductors [65].

**Acknowledgements**

This work is primarily supported by the US Department of Energy, Office of Science, Basic Energy Sciences, under Award Number DE-SC0020145 as part of the Computational Materials Sciences Program (V.G., L.-Q.C., and J.-M.H.). J.-M.H. also acknowledges the partial support from the National Science Foundation (NSF) under Grant No. DMR-2237884 on the dynamical phase-field simulations in this work. The dynamical phase-field simulations were performed using Bridges at the Pittsburgh Supercomputing Center through allocation TG-DMR180076 from the Advanced Cyberinfrastructure Coordination Ecosystem: Services & Support (ACCESS) program, which is supported by NSF Grants No. 2138259, No. 2138286, No. 2138307, No. 2137603, and No. 2138296. Part of this work was performed under the auspices of the U.S. Department of Energy by Lawrence Livermore National Laboratory under Contract DE-AC52-07NA27344 (B.W.). A.R acknowledges the support of the National Science Foundation Graduate Research Fellowship Program under Grant No. DGE1255832.

## Appendix A. Derivation of the equation of motion for the ionic polarization (soft mode) in ferroelectrics

We begin by writing the equation of motion for the soft mode, which has a coordinate $\mathbf{Q}^p$ ($Q_i^p$, $i=1,2,3$, with a unit of m) based on the Newton's second law, i.e.,

$$\mathbf{F}^{\text{ext}} + \mathbf{F}^{\text{restoring}} + \mathbf{F}^f = \rho^* \frac{d^2\mathbf{Q}^p}{dt^2}, \tag{A1}$$

where $\rho^*$ is the effective mass density of the soft mode (unit: kg/m³). $\mathbf{F}^{\text{ext}}$ is the volumetric coulombic force (unit: N/m³), given by,

$$F_i^{\text{ext}} = Z_P^*(E_i + E_i^d + E_i^{\text{rad}}), \tag{A2}$$

where $E_i$, $E_i^d$, and $E_i^{\text{rad}}$ are the electric field of the incident THz electromagnetic (EM) wave, the depolarization electric field, and the radiation electric field inside the ferroelectric, respectively, as mentioned in the main text. The subscript $i=x,y,z$ indicates that these electric fields are typically defined in the lab coordinate system, with $x\equiv1$, $y\equiv2$, $z\equiv3$ for the case of the (001)$_{pc}$ strained BaTiO$_3$ film and the (001)$_{pc}$ strained SrTiO$_3$ film (see Figs. 1(c) and 1(d)). The Born effective charge $Z_P^*$ describes the charge density associated with the soft mode. It has a unit of C/m³ and relates $\mathbf{Q}^p$ and ferroelectric polarization $\mathbf{P}^{\text{ion}}$ (or simply $\mathbf{P}$ hereafter) via $Q_i^p = P_i/Z_P^*$, as mentioned in the introduction.

The volumetric restoring force $F_i^{\text{restoring}}$, which is analogous to the restoring force that drives the electron back to its equilibrium position in nonlinear optical phenomena [26], is related to the local slope of the potential energy density landscape of the soft mode with contribution from both the Landau free energy density and elastic energy density. When the P is spatially uniform and oscillates in-phase in the ferroelectric (as is the case in the thin slab limit), $F_i^{\text{restoring}}$ is expressed as,

$$F_i^{\text{restoring}} = -\frac{\partial(f^{\text{Landau}} + f^{\text{Elast}})}{\partial Q_i^p} = -Z_P^* \frac{\partial(f^{\text{Landau}} + f^{\text{Elast}})}{\partial P_i} = Z_P^*(E_i^{\text{Landau}} + E_i^{\text{Elast}}), \tag{A3}$$

where the expressions of $f^{\text{Landau}}$ and $f^{\text{Elast}}$ are provided in Appendix B. Furthermore, the volumetric frictional force is assumed to be linearly proportional to the velocity of the soft mode, given by,

$$F_i^f = -\gamma_{ij}^* \frac{dQ_j^p}{dt} = -\frac{\gamma_{ij}^*}{Z_P^*} \frac{dP_j}{dt}. \tag{A4}$$

Substituting Eqs. (A2-4) into Eq. (A1) and using $P_i$ as the variable, one has, after some re-arrangement,

$$\frac{\rho^*}{Z_P^{*2}} \frac{d^2P_i}{dt^2} + \frac{\gamma_{ij}^*}{Z_P^{*2}} \frac{dP_j}{dt} = E_i + E_i^d + E_i^{\text{rad}} + E_i^{\text{Landau}} + E_i^{\text{Elast}}. \tag{A5}$$

By letting $\mu = \frac{\rho^*}{Z_P^{*2}}$, $\gamma_{ij} = \frac{\gamma_{ij}^*}{Z_P^{*2}}$, and using a total effective electric field $E_i^{\text{eff}}$ to represent the summation of all the fields on the right-hand side of Eq. (A5), the latter can be rewritten into a concise form, i.e.,

$$\mu \frac{d^2P_i}{dt^2} + \gamma_{ij} \frac{dP_j}{dt} = E_i^{\text{eff}}, \tag{A6}$$

which is equivalent to Eq. (3) because $P_i = P_i^0 + \Delta P_i$ and $E_i^{\text{eff}} = \Delta E_i^{\text{eff}}$, as discussed in the main text. Equation (3) can alternatively be derived based on an Euler-Lagrange approach by generalizing the procedures in [66] to consider the complete set of driving forces mentioned above.

**Appendix B. Detailed expressions of $f^\text{Landau}$ and $f^\text{Elast}$ for BaTiO₃, PbTiO₃, SrTiO₃, LiTaO₃ and LiNbO₃ and the associated materials parameters**

An eighth-order, sixth-order, and fourth-order $f^\text{Landau}$ is used for BaTiO₃, PbTiO₃, and SrTiO₃, respectively, given as [67],

$$\begin{aligned}f^\text{Landau} = &\alpha_1(P_1^2 + P_2^2 + P_3^2) + \alpha_{11}(P_1^4 + P_2^4 + P_3^4) + \alpha_{12}(P_1^2 P_2^2 + P_2^2 P_3^2 + P_1^2 P_3^2) \\ &+ \alpha_{111}(P_1^6 + P_2^6 + P_3^6) + \alpha_{112}[P_1^2(P_2^4 + P_3^4) + P_2^2(P_1^4 + P_3^4) + P_3^2(P_1^4 + P_2^4)] \\ &+ \alpha_{123} P_1^2 P_2^2 P_3^2 + \alpha_{1111}(P_1^8 + P_2^8 + P_3^8) \\ &+ \alpha_{1112}[P_1^6(P_2^2 + P_3^2) + P_2^6(P_1^2 + P_3^2) + P_3^6(P_1^2 + P_2^2)] + \alpha_{1122}(P_1^4 P_2^4 + P_2^4 P_3^4 + P_1^4 P_3^4) \\ &+ \alpha_{1123}(P_1^4 P_2^2 P_3^2 + P_2^4 P_1^2 P_3^2 + P_3^4 P_2^2 P_1^2)\end{aligned} \quad (B1)$$

$$\begin{aligned}f^\text{Landau} = &\alpha_1(P_1^2 + P_2^2 + P_3^2) + \alpha_{11}(P_1^4 + P_2^4 + P_3^4) + \alpha_{12}(P_1^2 P_2^2 + P_2^2 P_3^2 + P_1^2 P_3^2) \\ &+ \alpha_{111}(P_1^6 + P_2^6 + P_3^6) + \alpha_{112}[P_1^2(P_2^4 + P_3^4) + P_2^2(P_1^4 + P_3^4) + P_3^2(P_1^4 + P_2^4)] \\ &+ \alpha_{123} P_1^2 P_2^2 P_3^2\end{aligned} \quad (B2)$$

$$\begin{aligned}f^\text{Landau} = &\alpha_1(P_1^2 + P_2^2 + P_3^2) + \alpha_{11}(P_1^4 + P_2^4 + P_3^4) + \alpha_{12}(P_1^2 P_2^2 + P_2^2 P_3^2 + P_1^2 P_3^2) + \beta_1(q_1^2 + q_2^2 + q_3^2) \\ &+ \beta_{11}(q_1^4 + q_2^4 + q_3^4) + \beta_{12}(q_1^2 q_2^2 + q_2^2 q_3^2 + q_1^2 q_3^2) - t_{11}(P_1^2 q_1^2 + P_2^2 q_2^2 + P_3^2 q_3^2) \\ &- t_{12}(P_1^2(q_2^2 + q_3^2) + P_2^2(q_1^2 + q_3^2) + P_3^2(q_1^2 + q_2^2)) - t_{44}(P_1 P_2 q_1 q_2 + P_2 P_3 q_2 q_3 \\ &+ P_1 P_3 q_1 q_3)\end{aligned} \quad (B3)$$

Note that the $f^\text{Landau}$ of the SrTiO₃ also includes the terms that describe the coupling between the polarization $P_i$ the structural order parameter $q_i$ ($i$=1,2,3) which represents the linear oxygen displacement associated with oxygen octahedra rotation [68].

The elastic free energy density is given as $f^\text{Elast} = \frac{1}{2} c_{ijkl} e_{kl} e_{ij}$, where $e_{ij}$ and $e_{kl}$ are the elastic strains; $c_{ijkl}$ is the elastic stiffness tensor at constant electric field $E$ and temperature $T$, and can be related to the density of the electrical Helmholtz free energy $f$ (see its definition in Appendix D) via the relation $c_{ijkl} = \left(\frac{\partial \sigma_{ij}}{\partial e_{kl}}\right)_{T,E} = \left(\frac{\partial^2 f}{\partial e_{kl} \partial e_{ij}}\right)_{T,E}$. For the pseudocubic BaTiO₃, PbTiO₃, and SrTiO₃ films, $f^\text{Elast}$ is expanded as,

$$f^\text{Elast} = \frac{1}{2} c_{11}(e_{11}^2 + e_{22}^2 + e_{33}^2) + c_{12}(e_{11}e_{22} + e_{11}e_{33} + e_{22}e_{33}) + 2c_{44}(e_{12}^2 + e_{13}^2 + e_{23}^2), \quad (B4)$$

where the $c_{11}$, $c_{12}$, and $c_{44}$ are the independent components of the 6×6 elastic stiffness matrix $c_{mn}$ ($m$, $n$=1,2…6), which represents the elastic stiffness tensor $c_{ijkl}$ following the convention described in [69]. The $e_{ij} = \varepsilon_{ij} - \varepsilon_{ij}^0$ ($i,j$=1,2) is the elastic strain, where $\varepsilon_{ij}$ is the total strain and $\varepsilon_{ij}^0$ is the stress-free (eigen) strain. For BaTiO₃ and PbTiO₃, one has,

$$\varepsilon_{11}^0 = Q_{11} P_1^2 + Q_{12}(P_2^2 + P_3^2), \varepsilon_{22}^0 = Q_{11} P_2^2 + Q_{12}(P_1^2 + P_3^2), \varepsilon_{33}^0 = Q_{11} P_3^2 + Q_{12}(P_1^2 + P_2^2); \quad (B5a)$$

$$\varepsilon_{23}^0 = Q_{44} P_2 P_3, \varepsilon_{13}^0 = Q_{44} P_1 P_3, \varepsilon_{12}^0 = Q_{44} P_1 P_2, \quad (B5b)$$

where $Q_{11}$, $Q_{12}$, and $Q_{44}$ are the electrostrictive coefficients. For SrTiO₃, one has,

$$\varepsilon_{11}^0 = Q_{11} P_1^2 + Q_{12}(P_2^2 + P_3^2) + \Lambda_{11} q_1^2 + \Lambda_{12}(q_2^2 + q_3^2), \quad (B6a)$$

$$\varepsilon_{22}^0 = Q_{11} P_2^2 + Q_{12}(P_1^2 + P_3^2) + \Lambda_{11} q_2^2 + \Lambda_{12}(q_1^2 + q_3^2), \quad (B6b)$$

$$\varepsilon_{33}^0 = Q_{11} P_3^2 + Q_{12}(P_1^2 + P_2^2) + \Lambda_{11} q_3^2 + \Lambda_{12}(q_1^2 + q_2^2), \quad (B6c)$$

$$\varepsilon_{23}^0 = Q_{44}P_2P_3 + \Lambda_{44}q_2q_3, \varepsilon_{13}^0 = Q_{44}P_1P_3 + \Lambda_{44}q_1q_3, \varepsilon_{12}^0 = Q_{44}P_1P_2 + \Lambda_{44}q_1q_2, \quad \text{(B6d)}$$

where $\Lambda_{11}$, $\Lambda_{12}$, and $\Lambda_{44}$ are the linear quadratic coupling coefficient between the strain and structural order parameter. At the initial equilibrium state, the total strain $\varepsilon_{ij}$ is determined based on the mechanical boundary condition. In monodomain ($P_i$ and $q_i$ are spatially homogeneous), stress-free bulk crystals, $\varepsilon_{ij} = \varepsilon_{ij}^0$. In biaxially strained films, as indicated in Sect. II, $\varepsilon_{11} = \varepsilon_{11}^{mis}, \varepsilon_{22} = \varepsilon_{22}^{mis}, \varepsilon_{33} = -\frac{c_{12}}{c_{11}}(\varepsilon_{11}^{mis} + \varepsilon_{22}^{mis} - \varepsilon_{11}^0 - \varepsilon_{22}^0) + \varepsilon_{33}^0$, and $\varepsilon_{23} = \varepsilon_{23}^0, \varepsilon_{13} = \varepsilon_{13}^0, \varepsilon_{12} = 0$.

**Table B1**. List of the coefficients in the Landau and elastic free energy densities of BaTiO₃, PbTiO₃, and SrTiO₃. The temperature has a unit of °C for BaTiO₃ and PbTiO₃ and a unit of K for SrTiO₃.

| Coefficients | BaTiO₃ | PbTiO₃ | SrTiO₃ |
|---|---|---|---|
| $\alpha_1$ (N m² C⁻²) | $4.124 \times 10^5(T - 115)$ [70] | $3.8 \times 10^5(T - 479)$ [71] | $4.05 \times 10^7 \left(\coth\left(\frac{54}{T}\right) - \coth\left(\frac{54}{30}\right)\right)$ [72] |
| $\alpha_{11}$ (N m⁶ C⁻⁴) | $-2.097 \times 10^8$ [70] | $-0.73 \times 10^8$ [71] | $2.899 \times 10^9$ [31] |
| $\alpha_{12}$ (N m⁶ C⁻⁴) | $7.974 \times 10^8$ [70] | $7.5 \times 10^8$ [71] | $7.766 \times 10^9$ [68] |
| $\alpha_{111}$ (N m¹⁰ C⁻⁶) | $1.294 \times 10^9$ [70] | $2.6 \times 10^9$ [71] | 0 |
| $\alpha_{112}$ (N m¹⁰ C⁻⁶) | $-1.950 \times 10^9$ [70] | $6.1 \times 10^8$ [71] | 0 |
| $\alpha_{123}$ (N m¹⁰ C⁻⁶) | $-2.500 \times 10^9$ [70] | $-37 \times 10^8$ [71] | 0 |
| $\alpha_{1111}$ (N m¹⁴ C⁻⁸) | $3.863 \times 10^{10}$ [70] | 0 | 0 |
| $\alpha_{1112}$ (N m¹⁴ C⁻⁸) | $2.529 \times 10^{10}$ [70] | 0 | 0 |
| $\alpha_{1122}$ (N m¹⁴ C⁻⁸) | $1.637 \times 10^{10}$ [70] | 0 | 0 |
| $\alpha_{1123}$ (N m¹⁴ C⁻⁸) | $1.367 \times 10^{10}$ [70] | 0 | 0 |
| $\beta_1$ (N m⁻⁶) | 0 | 0 | $1.32 \times 10^{29}\left(\coth\left(\frac{145}{T}\right) - \coth\left(\frac{145}{105}\right)\right)$ [72] |
| $\beta_{11}$ (N m⁻⁶) | 0 | 0 | $1.688 \times 10^{50}$ [72] |
| $\beta_{12}$ (N m⁻⁶) | 0 | 0 | $3.879 \times 10^{50}$ [72] |
| $t_{11}$ (N m² C⁻²) | 0 | 0 | $-1.902 \times 10^{29}$ [31] |
| $t_{12}$ (N m² C⁻²) | 0 | 0 | $-1.014 \times 10^{29}$ [72] |
| $t_{44}$ (N m² C⁻²) | 0 | 0 | $5.865 \times 10^{29}$ [72] |
| $c_{11}$ (GPa) | 178 [70] | 174.6 [71] | 336 [72] |
| $c_{12}$ (GPa) | 96.4 [70] | 79.37 [71] | 107 [72] |
| $c_{44}$ (GPa) | 122 [70] | 111 [71] | 127 [72] |
| $Q_{11}$ (m⁴ C⁻²) | 0.1 [70] | 0.089 [71] | 0.0536 [72] |
| $Q_{12}$ (m⁴ C⁻²) | -0.034 [70] | -0.026 [71] | $-0.0154$ [72] |
| $Q_{44}$ (m⁴ C⁻²) | 0.029 [70] | 0.03375 [71] | 0.00472 [72] |
| $\Lambda_{11}$ (N C⁻²) | 0 | 0 | $8.820 \times 10^{18}$ [72] |
| $\Lambda_{12}$ (N C⁻²) | 0 | 0 | $-7.774 \times 10^{18}$ [72] |
| $\Lambda_{44}$ (N C⁻²) | 0 | 0 | $-4.528 \times 10^{18}$ [72] |

For trigonal crystals LiTaO₃ and LiNbO₃ that are uniaxial ferroelectrics, the Landau free energy density is written as [41],

$$f^{\text{Landau}} = -\frac{\alpha_1}{2}P_3^2 + \frac{\alpha_2}{4}P_3^4 + \frac{\alpha_3}{2}(P_1^2 + P_2^2), \quad \text{(B7)}$$

Following [41], the elastic energy density $f^{\text{Elas}}$ of the LiTaO₃ and LiNbO₃ is written, in contrast with Eqs. (B4-6), using Voigt notation, i.e.,

$$f^{\text{Elas}} = \beta_1\varepsilon_3^2 + \beta_2(\varepsilon_1+\varepsilon_2)^2 + \beta_3[(\varepsilon_1-\varepsilon_2)^2+\varepsilon_6^2] + \beta_4\varepsilon_3(\varepsilon_1+\varepsilon_2) + \beta_5(\varepsilon_4^2+\varepsilon_5^2)$$
$$+ \beta_6[(\varepsilon_1-\varepsilon_2)\varepsilon_4 + \varepsilon_5\varepsilon_6] + \gamma_1(\varepsilon_1+\varepsilon_2)P_3^2 + \gamma_2\varepsilon_3 P_3^2 + \gamma_3[(\varepsilon_1-\varepsilon_2)P_2P_3+\varepsilon_6 P_1 P_3]$$
$$+ \gamma_4(\varepsilon_5 P_1 P_3 + \varepsilon_4 P_2 P_3) + \gamma_5(\varepsilon_1+\varepsilon_2)(P_1^2+P_2^2) + \gamma_6\varepsilon_3(P_1^2+P_2^2)$$
$$+ \gamma_7[(\varepsilon_1-\varepsilon_2)(P_1^2-P_2^2)+2\varepsilon_6 P_1 P_2] + \gamma_8[\varepsilon_4(P_1^2-P_2^2)+2\varepsilon_5 P_1 P_2], \tag{B8}$$

where the $\varepsilon_i$ here are the total strain. For monodomain, stress-free bulk crystals, $\varepsilon_i$ can be obtained by solving the mechanical boundary condition $\sigma_j = \partial f^{\text{Elas}}/\partial \varepsilon_j = 0$, $j=1,2,3,4,5,6$, given by,

$$\varepsilon_1 = \Gamma_1 P_1^2 + \Gamma_2 P_2^2 + \Gamma_3 P_3^2 + \Gamma_4 P_2 P_3, \quad \varepsilon_2 = \Gamma_2 P_1^2 + \Gamma_1 P_2^2 + \Gamma_3 P_3^2 - \Gamma_4 P_2 P_3, \tag{B9a}$$

$$\varepsilon_3 = \Gamma_5 P_1^2 + \Gamma_5 P_2^2 + \Gamma_6 P_3^2, \quad \varepsilon_4 = \Gamma_7 P_1^2 - \Gamma_7 P_2^2 + \Gamma_8 P_2 P_3, \tag{B9b}$$

$$\varepsilon_5 = \Gamma_8 P_1 P_3 + 2\Gamma_7 P_1 P_2, \quad \varepsilon_6 = 2\Gamma_4 P_1 P_3 + \Gamma_9 P_1 P_2, \tag{B9c}$$

Here the coefficients $\Gamma_i$ are $\Gamma_1 = \frac{\beta_1(-8\beta_3\beta_5\gamma_5+2\beta_6^2\gamma_5-8\beta_2\beta_5\gamma_7+4\beta_2\beta_6\gamma_8)+\beta_4(4\beta_3\beta_5\gamma_6-\beta_6^2\gamma_6+2\beta_4\beta_5\gamma_7-\beta_4\beta_6\gamma_8)}{2(4\beta_1\beta_2-\beta_4^2)(4\beta_3\beta_5-\beta_6^2)}$,
$\Gamma_2 = \frac{\beta_1(-8\beta_3\beta_5\gamma_5+2\beta_6^2\gamma_5+8\beta_2\beta_5\gamma_7-4\beta_2\beta_6\gamma_8)+\beta_4(4\beta_3\beta_5\gamma_6-\beta_6^2\gamma_6-2\beta_4\beta_5\gamma_7+\beta_4\beta_6\gamma_8)}{2(4\beta_1\beta_2-\beta_4^2)(4\beta_3\beta_5-\beta_6^2)}$, $\Gamma_3 = \frac{-2\beta_1\gamma_1+\beta_4\gamma_2}{2(4\beta_1\beta_2-\beta_4^2)}$, $\Gamma_4 = \frac{-2\beta_5\gamma_3+\beta_6\gamma_4}{2(4\beta_3\beta_5-\beta_6^2)}$, $\Gamma_5 = \frac{\beta_4\gamma_5-2\beta_2\gamma_6}{4\beta_1\beta_2-\beta_4^2}$, $\Gamma_6 = \frac{\beta_4\gamma_1-2\beta_2\gamma_2}{4\beta_1\beta_2-\beta_4^2}$, $\Gamma_7 = \frac{\beta_6\gamma_7-2\beta_3\gamma_8}{4\beta_3\beta_5-\beta_6^2}$, $\Gamma_8 = \frac{\beta_6\gamma_3-2\beta_3\gamma_4}{4\beta_3\beta_5-\beta_6^2}$, $\Gamma_9 = \frac{2\beta_6\gamma_8-4\beta_5\gamma_7}{4\beta_3\beta_5-\beta_6^2}$.

**Table B2**. List of the coefficients in the Landau and elastic free energy densities of LiTaO$_3$ and LiNbO$_3$

| Coefficients | LiTaO$_3$ | LiNbO$_3$ |
|---|---|---|
| $\alpha_1$ (N m$^2$ C$^{-2}$) | $1.25 \times 10^9$ (fitted from [39]) | $2.012 \times 10^9$ [41] |
| $\alpha_2$ (N m$^2$ C$^{-2}$) | $6 \times 10^8$ (fitted from [39]) | $3.608 \times 10^9$ [41] |
| $\alpha_3$ (N m$^2$ C$^{-2}$) | $1.3 \times 10^9$ (fitted from [39]) | $1.345 \times 10^9$ [41] |
| $\beta_1$ (N m$^{-2}$) | $13.55 \times 10^{10}$ [41] | $12.25 \times 10^{10}$ [41] |
| $\beta_2$ (N m$^{-2}$) | $6.475 \times 10^{10}$ [41] | $6.4 \times 10^{10}$ [41] |
| $\beta_3$ (N m$^{-2}$) | $4.925 \times 10^{10}$ [41] | $3.75 \times 10^{10}$ [41] |
| $\beta_4$ (N m$^{-2}$) | $7.4 \times 10^{10}$ [41] | $7.5 \times 10^{10}$ [41] |
| $\beta_5$ (N m$^{-2}$) | $4.8 \times 10^{10}$ [41] | $3 \times 10^{10}$ [41] |
| $\beta_6$ (N m$^{-2}$) | $-1.2 \times 10^{10}$ [41] | $0.9 \times 10^{10}$ [41] |
| $\gamma_1$ (N m$^2$ C$^{-2}$) | $-0.202 \times 10^9$ [41] | $0.216 \times 10^9$ [41] |
| $\gamma_2$ (N m$^2$ C$^{-2}$) | $1.317 \times 10^9$ [41] | $1.848 \times 10^9$ [41] |
| $\gamma_3$ (N m$^2$ C$^{-2}$) | $-2.824 \times 10^9$ [41] | $-0.33 \times 10^9$ [41] |
| $\gamma_4$ (N m$^2$ C$^{-2}$) | $4.992 \times 10^9$ [41] | $3.9 \times 10^9$ [41] |

Table B1 and B2 list the coefficients in the Landau and elastic free energy densities of the four ferroelectric materials. The mass coefficient $\mu = 1.35 \times 10^{-18}$ J m s$^2$ C$^{-2}$ for BaTiO$_3$ [73], $\mu = 1.59 \times 10^{-18}$ J m s$^2$ C$^{-2}$ for PbTiO$_3$ [73], $\mu = 22 \times 10^{-18}$ J m s$^2$ C$^{-2}$ for SrTiO$_3$ [74], $\mu = 1.81 \times 10^{-18}$ J m s$^2$ C$^{-2}$ for LiNbO$_3$ [29]. The $\mu$ value of the LiTaO$_3$, which is not yet available in literature to our knowledge, is set to be the same as the LiNbO$_3$.

## Appendix C. Linear and nonlinear susceptibility derivation by perturbation method

Given that $E_i^{\text{Landau}}(t) = E_i^{\text{Landau}}(P_i^0) + \Delta E_i^{\text{Landau}}(t)$, $E_i^{\text{Elast}}(t) = E_i^{\text{Elast}}(P_i^0) + \Delta E_i^{\text{Elast}}(t)$, $E_i^{\text{d}} = E_i^{\text{d}}(P_i^0) + \Delta E_i^{\text{d}}(t)$, and that $E_i^{\text{Landau}}(P_i^0) + E_i^{\text{Elast}}(P_i^0) + E_i^{\text{d}}(P_i^0) = 0$ in the initial equilibrium state ($P_i = P_i^0$), the equation of motion for polarization (Eq. (3)) can be rewritten as,

$$\mu \frac{\partial^2 \Delta P_i}{\partial t^2} + \gamma_{ij} \frac{\partial \Delta P_j}{\partial t} = \Delta E_i^{\text{Landau}} + \Delta E_i^{\text{Elast}} + \Delta E_i^{\text{d}} + E_i + E_i^{\text{rad}}, \tag{C1}$$

where $\Delta E_i^{\text{Landau}}$ and $\Delta E_i^{\text{Elast}}$ can be expanded through Taylor series expansion, i.e., $\Delta E_i^{\text{Landau}} = A_{ij} \Delta P_j + \frac{1}{2} A_{ijk} \Delta P_j \Delta P_k + \cdots$, $\Delta E_i^{\text{Elast}} = B_{ij} \Delta P_j + \frac{1}{2} B_{ijk} \Delta P_j \Delta P_k + \cdots$, where the expressions of $A_{ij}$, $A_{ijk}$, $B_{ij}$, and $B_{ijk}$ are provided in Sec. II, with $i, j = 1,2,3$; $\Delta E_i^{\text{d}} = (0, 0, -\frac{1}{\kappa_0 \kappa_b} \Delta P_z)$ and $E_i^{\text{rad}} = (-\frac{d_0}{2\kappa_0 c} \frac{\partial P_x}{\partial t}, -\frac{d_0}{2\kappa_0 c} \frac{\partial P_y}{\partial t}, 0)$ are often expressed in the $x$-$y$-$z$ lab coordinate system, with $x \equiv 1$, $y \equiv 2$, and $z \equiv 3$. Moreover, we assume that only the diagonal components of the 3×3 matrix $\gamma_{ij}$ are non-zero. This assumption is also adopted in the derivation of the nonlinear susceptibility of electronic polarization to optical light waves [26]. The values of $\gamma_{11}$, $\gamma_{22}$, and $\gamma_{33}$ determine the linewidth of a peak in the frequency spectra of both the linear and nonlinear susceptibilities, and therefore can be calibrated by comparing the theoretically predicted and experimentally measured linewidth. If only keeping the first two terms in the expanded expressions of $\Delta E_i^{\text{Landau}}$ and $\Delta E_i^{\text{Elast}}$ and Eq. (C1) can be expanded into the following equations,

$$\mu \frac{\partial^2 \Delta P_1}{\partial t^2} + (\gamma_{11} + \frac{d_0}{2\kappa_0 c}) \frac{\partial \Delta P_1}{\partial t} - (A_{11}+B_{11})\Delta P_1 - (A_{12}+B_{12})\Delta P_2 - (A_{13}+B_{13})\Delta P_3 - \frac{1}{2}(A_{111}+B_{111})\Delta P_1^2 - \frac{1}{2}(A_{122}+B_{122})\Delta P_1^2 - \frac{1}{2}(A_{133}+B_{133})\Delta P_1^2 - (A_{123}+B_{123})\Delta P_2 \Delta P_3 - (A_{113}+B_{113})\Delta P_1 \Delta P_3 - (A_{112}+B_{112})\Delta P_1 \Delta P_2 = E_1, \tag{C2a}$$

$$\mu \frac{\partial^2 \Delta P_2}{\partial t^2} + (\gamma_{22} + \frac{d_0}{2\kappa_0 c}) \frac{\partial \Delta P_2}{\partial t} - (A_{21}+B_{21})\Delta P_1 - (A_{22}+B_{22})\Delta P_2 - (A_{23}+B_{23})\Delta P_3 - \frac{1}{2}(A_{211}+B_{211})\Delta P_1^2 - \frac{1}{2}(A_{222}+B_{222})\Delta P_1^2 - \frac{1}{2}(A_{233}+B_{133})\Delta P_1^2 - (A_{223}+B_{223})\Delta P_2 \Delta P_3 - (A_{213}+B_{213})\Delta P_1 \Delta P_3 - (A_{212}+B_{212})\Delta P_1 \Delta P_2 = E_2, \tag{C2b}$$

$$\mu \frac{\partial^2 \Delta P_3}{\partial t^2} + \gamma_{33} \frac{\partial \Delta P_3}{\partial t} - (A_{31}+B_{31})\Delta P_1 - (A_{32}+B_{32})\Delta P_2 - (A_{33}+B_{33}-\frac{1}{\kappa_0 \kappa_b})\Delta P_3 - \frac{1}{2}(A_{311}+B_{311})\Delta P_1^2 - \frac{1}{2}(A_{322}+B_{322})\Delta P_1^2 - \frac{1}{2}(A_{333}+B_{333})\Delta P_1^2 - (A_{323}+B_{323})\Delta P_2 \Delta P_3 - (A_{313}+B_{313})\Delta P_1 \Delta P_3 - (A_{312}+B_{312})\Delta P_1 \Delta P_2 = E_3, \tag{C2c}$$

Equations C2(a-c) can be written into a matrix form as shown by Eq. (4), reproduced below,

$$\mu \frac{\partial^2 \Delta \mathbf{P}}{\partial t^2} + \boldsymbol{\gamma}^{\text{eff}} \frac{\partial \Delta \mathbf{P}}{\partial t} + \mathbf{K} \Delta \mathbf{P} + \mathbf{C} \Delta \mathbf{P}_{\text{II}} = \mathbf{E}, \tag{C3}$$

where the expanded expressions of the matrices $\Delta \mathbf{P}$, $\mathbf{E}$, $\Delta \mathbf{P}_{\text{II}}$, and $\boldsymbol{\gamma}^{\text{eff}}$ as well as the tensors $\mathbf{K}$ and $\mathbf{C}$ are provided in the main paper, as shown by Eqs. (4), 6(a), 6(b) and related text. To solve Eq. (C3) by the perturbation method, we begin by replacing $\mathbf{E}$ with $\lambda \mathbf{E}$, where $\lambda$ is a parameter that characterizes the strength of the perturbation and ranges continuously between zero and one and will be set equal to one at the end of calculation. In the framework of perturbation method, the solution of Eq. (C3) can be written as $\Delta \mathbf{P} = \lambda \Delta \mathbf{P}^{(1)} + \lambda^2 \Delta \mathbf{P}^{(2)} + \lambda^3 \Delta \mathbf{P}^{(3)} + \cdots$, where $\Delta \mathbf{P}^{(1)}$ is the lowest-order (linear) contribution to the $\Delta \mathbf{P}$, calculated as $\Delta \mathbf{P}^{(1)} = \Delta P_i^{\text{Linear}} = \kappa_0 \chi_{ij}^{(1)} E_j$, $\Delta \mathbf{P}^{(2)}$ is the second-order nonlinear term of nonlinear polarization oscillation, calculated as $\Delta \mathbf{P}^{(2)} = \kappa_0 \chi_{ijk}^{(2)} E_j E_k$, and so forth.

It is worth emphasizing that the premise of the perturbation theory in the present application is that the center of polarization oscillation is always at $P_i = P_i^0$, which is only valid when the amplitude of the $E_i^{inc}$ is not too large. Under strong excitation, the d.c. polarization shift $\Delta P_i^{(2)}(0)$, as discussed in the text after Eq. (6), would be large and therefore shifts the center of polarization oscillation from $P_i^0$ to $P_i^0 + \Delta P_i^{(2)}(0)$. Alternatively, the polarization dynamics under strong excitation can be obtained by numerical solutions from a dynamical phase-field model with coupled strain-polarization-EM wave dynamics [18,25]. With this in mind, we proceed by writing the term $\Delta \mathbf{P}_{II}$ as:

$$\Delta \mathbf{P}_{II} = \left(\left(\sum_{i=1}^{2}\lambda^i \Delta P_1^{(i)}\right)^2, \left(\sum_{i=1}^{2}\lambda^i \Delta P_2^{(i)}\right)^2, \left(\sum_{i=1}^{2}\lambda^i \Delta P_3^{(i)}\right)^2,\right.$$

$$\left. 2\left(\sum_{i=1}^{2}\lambda^i \Delta P_2^{(i)}\right)\left(\sum_{i=1}^{2}\lambda^i \Delta P_3^{(i)}\right), 2\left(\sum_{i=1}^{2}\lambda^i \Delta P_1^{(i)}\right)\left(\sum_{i=1}^{2}\lambda^i \Delta P_3^{(i)}\right), 2\left(\sum_{i=1}^{2}\lambda^i \Delta P_1^{(i)}\right)\left(\sum_{i=1}^{2}\lambda^i \Delta P_2^{(i)}\right)\right)^T$$

$$= \lambda^2 [\Delta P_1^{(1)^2}, \Delta P_2^{(1)^2}, \Delta P_3^{(1)^2}, 2\Delta P_2^{(1)}\Delta P_3^{(1)}, 2\Delta P_1^{(1)}\Delta P_3^{(1)}, 2\Delta P_1^{(1)}\Delta P_2^{(1)}]^T$$

$$+ \lambda^3 [2\Delta P_1^{(1)}\Delta P_1^{(2)}, 2\Delta P_2^{(1)}\Delta P_2^{(2)}, 2\Delta P_3^{(1)}\Delta P_3^{(2)},$$

$$2\Delta P_2^{(1)}\Delta P_3^{(2)} + 2\Delta P_3^{(1)}\Delta P_2^{(2)}, 2\Delta P_1^{(1)}\Delta P_3^{(2)} + 2\Delta P_3^{(1)}\Delta P_1^{(2)}, 2\Delta P_1^{(1)}\Delta P_2^{(2)} + 2\Delta P_2^{(1)}\Delta P_1^{(2)}]^T + \cdots$$

$$= \lambda^2 \Delta \mathbf{P}_{II}^{(1)} + \lambda^3 \Delta \mathbf{P}_{II}^{(1,2)} + \cdots \quad (C4)$$

Plugging in the expression $\Delta \mathbf{P} = \lambda \Delta \mathbf{P}^{(1)} + \lambda^2 \Delta \mathbf{P}^{(2)} + \lambda^3 \Delta \mathbf{P}^{(3)} + \cdots$ and Eq. (C4) into Eq. (C3), Eq.(C3) can be rewritten into a form given by $\lambda(\text{Eq. }u_1) + \lambda^2(\text{Eq. }u_2) + \lambda^3(\text{Eq. }u_3) + \cdots = \lambda \mathbf{E}^{inc}$. The solution to this equation requires that Eq. $u_1 = \mathbf{E}^{inc}$ and that Eq. $u_n = 0$ (n=2,3,4…), which can be expanded into a series of linear equations as follows,

$$\mu \frac{\partial^2 \Delta \mathbf{P}^{(1)}}{\partial t^2} + \mathbf{\gamma}^{eff} \frac{\partial \Delta \mathbf{P}^{(1)}}{\partial t} + \mathbf{K}\Delta \mathbf{P}^{(1)} = \mathbf{E}^{inc} \quad (C5a)$$

$$\mu \frac{\partial^2 \Delta \mathbf{P}^{(2)}}{\partial t^2} + \mathbf{\gamma}^{eff} \frac{\partial \Delta \mathbf{P}^{(2)}}{\partial t} + \mathbf{K}\Delta \mathbf{P}^{(2)} + \mathbf{C}\Delta \mathbf{P}_{II}^{(1)} = 0 \quad (C5b)$$

$$\mu \frac{\partial^2 \Delta \mathbf{P}^{(3)}}{\partial t^2} + \mathbf{\gamma}^{eff} \frac{\partial \Delta \mathbf{P}^{(3)}}{\partial t} + \mathbf{K}\Delta \mathbf{P}^{(3)} + \mathbf{C}\Delta \mathbf{P}_{II}^{(1,2)} = 0 \quad (C5c)$$

Under a single-frequency continuous incident THz wave in the thin slab limit, $E_i^{inc}(t) = E_i^{inc,0}e^{-i\omega t}$, one can write $\Delta \mathbf{P}^{(1)} = \Delta \mathbf{P}^0 e^{i(-\omega t + \varphi)}$ as a steady-state solution, where the $\varphi$ is the phase difference between the incident THz wave and the excited polarization wave. Accordingly, Eq. (C5a) can be rewritten as,

$$-\mu\omega^2 \Delta \mathbf{P}^{(1)} - i\mathbf{\gamma}^{eff}\omega \Delta \mathbf{P}^{(1)} + \mathbf{K}\Delta \mathbf{P}^{(1)} = \mathbf{E}^{inc}. \quad (C6)$$

Rearranging Eq. (C6), one has $\Delta \mathbf{P}^{(1)} = (\mathbf{K} - \mu\omega^2 - i\mathbf{\gamma}^{eff}\omega)^{-1}\mathbf{E}^{inc} = \kappa_0 \chi_{ij}^{(1)} E_j^{inc}$, from which the expression of $\chi_{ij}^{(1)}$ can be derived, as shown by Eq. (4) in Sect. II.

Substituting the steady-state solution of $\Delta \mathbf{P}^{(1)}$ into Eq. (C5b) allows for deriving the steady-state solution of $\Delta \mathbf{P}^{(2)}$ and therefore the $\chi^{(2)}_{ijk}$. To do this, we first expand the terms that are contained in the expression $\Delta \mathbf{P}^{(1)}_{\mathrm{II}}$ as follows,

$$\Delta P^{(1)}_j(\omega)\Delta P^{(1)}_k(\omega) = \kappa_0^2 |\chi_{jm}| E^{\mathrm{inc},0}_m \cos(\omega t + \varphi^{jm}) |\chi_{kn}| E^{\mathrm{inc},0}_n \cos(\omega t + \varphi^{kn})$$

$$= \kappa_0^2 \big(|\chi_{jm}| E^{\mathrm{inc},0}_m\big)\big(|\chi_{kn}| E^{\mathrm{inc},0}_n\big)\left(\frac{1}{2}\cos(\varphi^{jm} - \varphi^{kn}) + \frac{1}{2}\cos(2\omega t + \varphi^{jm} + \varphi^{kn})\right)$$

$$\equiv \frac{\kappa_0^2}{2}\chi^{(1)}_{jm}\chi^{(1),*}_{kn} E^{\mathrm{inc},0}_m E^{\mathrm{inc},0}_n + \frac{\kappa_0^2}{2}\chi^{(1)}_{jm}\chi^{(1)}_{kn} E^{\mathrm{inc},0}_m E^{\mathrm{inc},0}_n e^{-\mathrm{i}2\omega t} = \Delta P^{(2)}(0) + \Delta P^{(2)}(2\omega), \qquad (C7)$$

which therefore contains both a d.c. shift $\Delta \mathbf{P}^{(2)}(0)$ and a second-order harmonic component $\Delta \mathbf{P}^{(2)}(2\omega)$. Here $\varphi^{jm}$ refers to the phase difference between the oscillatory polarization component $\Delta P^{(1)}_j(t)$ and the excitation electric field component $E^{\mathrm{inc}}_m$, and so forth for the $\varphi^{kn}$. $\chi^{(1),*}_{kn} = \chi^{(1),\mathrm{Re}}_{kn} - \mathrm{i}\chi^{(1),\mathrm{Im}}_{kn}$ is the conjugation of the complex susceptibility of $\chi^{(1)}_{kn} = \chi^{(1),\mathrm{Re}}_{kn} + \mathrm{i}\chi^{(1),\mathrm{Im}}_{kn}$. Plugging both the steady-state solution of $\Delta \mathbf{P}^{(1)}$ and Eq. (C7) into Eq. (C5b), and if only considering the $\Delta \mathbf{P}^{(2)}(2\omega)$, Eq. (C5b) can be rewritten as,

$$-\mu 4\omega^2 \Delta \mathbf{P}^{(2)} - \mathrm{i}\gamma^{\mathrm{eff}}_{ii} 2\omega \Delta \mathbf{P}^{(2)} + \mathbf{K}\Delta \mathbf{P}^{(2)} = -\frac{\kappa_0^2}{2}\sum_{j,k=1,2,3} C_{ijk} \left(\sum_{j,m=1,2,3}\chi^{(1)}_{jm} E^{\mathrm{inc},0}_m\right)\left(\sum_{k,n=1,2,3}\chi^{(1)}_{kn} E^{\mathrm{inc},0}_n\right). \quad (C8)$$

Thus, the SHG susceptibility $\chi^{(2)}_{ijk}(2\omega)$ can be expanded as,

$$\chi^{(2)}_{ijk}(2\omega,\omega,\omega) = -\kappa_0^2 \sum_{\alpha,\beta,\gamma=1,2,3} C_{\alpha\beta\gamma}\chi^{(1)}_{i\alpha}(2\omega)\chi^{(1)}_{\beta j}(\omega)\chi^{(1)}_{\gamma k}(\omega)\;,\; i,j,k=1,2,3$$

$$= \begin{bmatrix} \chi^{(2)}_{111} & \chi^{(2)}_{122} & \chi^{(2)}_{133} & \chi^{(2)}_{123} & \chi^{(2)}_{113} & \chi^{(2)}_{112} \\ \chi^{(2)}_{211} & \chi^{(2)}_{222} & \chi^{(2)}_{233} & \chi^{(2)}_{223} & \chi^{(2)}_{213} & \chi^{(2)}_{212} \\ \chi^{(2)}_{311} & \chi^{(2)}_{322} & \chi^{(2)}_{333} & \chi^{(2)}_{323} & \chi^{(2)}_{313} & \chi^{(2)}_{312} \end{bmatrix}$$

$$= -\kappa_0^2 \begin{bmatrix} \chi^{(1)}_{11}(2\omega) & \chi^{(1)}_{12}(2\omega) & \chi^{(1)}_{13}(2\omega) \\ \chi^{(1)}_{21}(2\omega) & \chi^{(1)}_{22}(2\omega) & \chi^{(1)}_{23}(2\omega) \\ \chi^{(1)}_{31}(2\omega) & \chi^{(1)}_{32}(2\omega) & \chi^{(1)}_{33}(2\omega) \end{bmatrix} \cdot \begin{bmatrix} C_{111} & C_{122} & C_{133} & C_{123} & C_{113} & C_{112} \\ C_{211} & C_{222} & C_{233} & C_{223} & C_{213} & C_{212} \\ C_{311} & C_{322} & C_{333} & C_{323} & C_{313} & C_{312} \end{bmatrix} \cdot$$

$$\begin{bmatrix} \chi^{(1)^2}_{11}(\omega) & \chi^{(1)^2}_{12}(\omega) & \chi^{(1)^2}_{13}(\omega) & 2\chi^{(1)}_{12}(\omega)\chi^{(1)}_{13}(\omega) \\ \chi^{(1)^2}_{21}(\omega) & \chi^{(1)^2}_{22}(\omega) & \chi^{(1)^2}_{23}(\omega) & 2\chi^{(1)}_{22}(\omega)\chi^{(1)}_{23}(\omega) \\ \chi^{(1)^2}_{31}(\omega) & \chi^{(1)^2}_{32}(\omega) & \chi^{(1)^2}_{33}(\omega) & 2\chi^{(1)}_{32}(\omega)\chi^{(1)}_{33}(\omega) \\ 2\chi^{(1)}_{21}(\omega)\chi^{(1)}_{31}(\omega) & 2\chi^{(1)}_{22}(\omega)\chi^{(1)}_{32}(\omega) & 2\chi^{(1)}_{23}(\omega)\chi^{(1)}_{33}(\omega) & 2\chi^{(1)}_{22}(\omega)\chi^{(1)}_{33}(\omega) + 2\chi^{(1)}_{23}(\omega)\chi^{(1)}_{32}(\omega) \\ 2\chi^{(1)}_{11}(\omega)\chi^{(1)}_{31}(\omega) & 2\chi^{(1)}_{12}(\omega)\chi^{(1)}_{32}(\omega) & 2\chi^{(1)}_{13}(\omega)\chi^{(1)}_{33}(\omega) & 2\chi^{(1)}_{12}(\omega)\chi^{(1)}_{33}(\omega) + 2\chi^{(1)}_{13}(\omega)\chi^{(1)}_{32}(\omega) \\ 2\chi^{(1)}_{11}(\omega)\chi^{(1)}_{21}(\omega) & 2\chi^{(1)}_{12}(\omega)\chi^{(1)}_{22}(\omega) & 2\chi^{(1)}_{13}(\omega)\chi^{(1)}_{23}(\omega) & 2\chi^{(1)}_{12}(\omega)\chi^{(1)}_{23}(\omega) + 2\chi^{(1)}_{13}(\omega)\chi^{(1)}_{22}(\omega) \end{bmatrix}$$

$$\begin{matrix} 2\chi_{11}^{(1)}(\omega)\chi_{13}^{(1)}(\omega) & 2\chi_{11}^{(1)}(\omega)\chi_{12}^{(1)}(\omega) \\ 2\chi_{21}^{(1)}(\omega)\chi_{23}^{(1)}(\omega) & 2\chi_{21}^{(1)}(\omega)\chi_{22}^{(1)}(\omega) \\ 2\chi_{31}^{(1)}(\omega)\chi_{33}^{(1)}(\omega) & 2\chi_{31}^{(1)}(\omega)\chi_{32}^{(1)}(\omega) \\ 2\chi_{21}^{(1)}(\omega)\chi_{33}^{(1)}(\omega) + 2\chi_{23}^{(1)}(\omega)\chi_{31}^{(1)}(\omega) & 2\chi_{21}^{(1)}(\omega)\chi_{32}^{(1)}(\omega) + 2\chi_{22}^{(1)}(\omega)\chi_{31}^{(1)}(\omega) \\ 2\chi_{11}^{(1)}(\omega)\chi_{33}^{(1)}(\omega) + 2\chi_{13}^{(1)}(\omega)\chi_{31}^{(1)}(\omega) & 2\chi_{11}^{(1)}(\omega)\chi_{32}^{(1)}(\omega) + 2\chi_{12}^{(1)}(\omega)\chi_{31}^{(1)}(\omega) \\ 2\chi_{11}^{(1)}(\omega)\chi_{23}^{(1)}(\omega) + 2\chi_{13}^{(1)}(\omega)\chi_{21}^{(1)}(\omega) & 2\chi_{11}^{(1)}(\omega)\chi_{22}^{(1)}(\omega) + 2\chi_{12}^{(1)}(\omega)\chi_{21}^{(1)}(\omega) \end{matrix} \Bigg] \quad (C9)$$

The d.c. shift component $\Delta \mathbf{P}^{(2)}(0)$ can be derived from the following equation,

$$\mathbf{K}\Delta \mathbf{P}^{(2)}(0) = -\frac{\kappa_0^2}{2} \sum_{j,k=1,2,3} C_{ijk} \left( \sum_{j,m=1,2,3} \chi_{jm}^{(1)} E_m^{inc,0} \right) \left( \sum_{k,n=1,2,3} \chi_{kn}^{(1),*} E_n^{inc,0} \right). \quad (C10)$$

Accordingly, the nonlinear electric susceptibility $\chi_{ijk}^{(2)}(0)$ can be derived as,

$$\chi_{ijk}^{(2)}(0,\omega,-\omega) = -\kappa_0^2 \sum_{\alpha,\beta,\gamma=1,2,3} C_{\alpha\beta\gamma} \chi_{i\alpha}^{(1)}(0)\chi_{\beta j}^{(1)}(\omega)\chi_{\gamma k}^{(1),*}(\omega), i,j,k = 1,2,3$$

$$= \begin{bmatrix} \chi_{111}^{(2)} & \chi_{122}^{(2)} & \chi_{133}^{(2)} & \chi_{123}^{(2)} & \chi_{113}^{(2)} & \chi_{112}^{(2)} \\ \chi_{211}^{(2)} & \chi_{222}^{(2)} & \chi_{233}^{(2)} & \chi_{223}^{(2)} & \chi_{213}^{(2)} & \chi_{212}^{(2)} \\ \chi_{311}^{(2)} & \chi_{322}^{(2)} & \chi_{333}^{(2)} & \chi_{323}^{(2)} & \chi_{313}^{(2)} & \chi_{312}^{(2)} \end{bmatrix}$$

$$= -\kappa_0^2 \begin{bmatrix} \chi_{11}^{(1)}(0) & \chi_{12}^{(1)}(0) & \chi_{13}^{(1)}(0) \\ \chi_{21}^{(1)}(0) & \chi_{22}^{(1)}(0) & \chi_{23}^{(1)}(0) \\ \chi_{31}^{(1)}(0) & \chi_{32}^{(1)}(0) & \chi_{33}^{(1)}(0) \end{bmatrix} \cdot \begin{bmatrix} C_{111} & C_{122} & C_{133} & C_{123} & C_{113} & C_{112} \\ C_{211} & C_{222} & C_{233} & C_{223} & C_{213} & C_{212} \\ C_{311} & C_{322} & C_{333} & C_{323} & C_{313} & C_{312} \end{bmatrix} \cdot$$

$$\begin{bmatrix} |\chi_{11}^{(1)}(\omega)|^2 & |\chi_{12}^{(1)}(\omega)|^2 & |\chi_{13}^{(1)}(\omega)|^2 & 2\chi_{12}^{(1)}(\omega)\chi_{13}^{(1),*}(\omega) \\ |\chi_{21}^{(1)}(\omega)|^2 & |\chi_{22}^{(1)}(\omega)|^2 & |\chi_{23}^{(1)}(\omega)|^2 & 2\chi_{22}^{(1)}(\omega)\chi_{23}^{(1),*}(\omega) \\ |\chi_{31}^{(1)}(\omega)|^2 & |\chi_{32}^{(1)}(\omega)|^2 & |\chi_{33}^{(1)}(\omega)|^2 & 2\chi_{32}^{(1)}(\omega)\chi_{33}^{(1),*}(\omega) \\ 2\chi_{21}^{(1)}(\omega)\chi_{31}^{(1),*}(\omega) & 2\chi_{22}^{(1)}(\omega)\chi_{32}^{(1),*}(\omega) & 2\chi_{23}^{(1)}(\omega)\chi_{33}^{(1),*}(\omega) & 2\chi_{22}^{(1)}(\omega)\chi_{33}^{(1),*}(\omega) + 2\chi_{23}^{(1)}(\omega)\chi_{32}^{(1),*}(\omega) \\ 2\chi_{11}^{(1)}(\omega)\chi_{31}^{(1),*}(\omega) & 2\chi_{12}^{(1)}(\omega)\chi_{32}^{(1),*}(\omega) & 2\chi_{13}^{(1)}(\omega)\chi_{33}^{(1),*}(\omega) & 2\chi_{12}^{(1)}(\omega)\chi_{33}^{(1),*}(\omega) + 2\chi_{13}^{(1)}(\omega)\chi_{32}^{(1),*}(\omega) \\ 2\chi_{11}^{(1)}(\omega)\chi_{21}^{(1),*}(\omega) & 2\chi_{12}^{(1)}(\omega)\chi_{22}^{(1),*}(\omega) & 2\chi_{13}^{(1)}(\omega)\chi_{23}^{(1),*}(\omega) & 2\chi_{12}^{(1)}(\omega)\chi_{23}^{(1),*}(\omega) + 2\chi_{13}^{(1)}(\omega)\chi_{22}^{(1),*}(\omega) \end{bmatrix}$$

$$\begin{matrix} 2\chi_{11}^{(1)}(\omega)\chi_{13}^{(1),*}(\omega) & 2\chi_{11}^{(1)}(\omega)\chi_{12}^{(1),*}(\omega)) \\ 2\chi_{21}^{(1)}(\omega)\chi_{23}^{(1),*}(\omega) & 2\chi_{21}^{(1)}(\omega)\chi_{22}^{(1),*}(\omega) \\ 2\chi_{31}^{(1)}(\omega)\chi_{33}^{(1),*}(\omega) & 2\chi_{31}^{(1)}(\omega)\chi_{32}^{(1),*}(\omega) \\ 2\chi_{21}^{(1)}(\omega)\chi_{33}^{(1),*}(\omega) + 2\chi_{23}^{(1)}(\omega)\chi_{31}^{(1),*}(\omega) & 2\chi_{21}^{(1)}(\omega)\chi_{32}^{(1),*}(\omega) + 2\chi_{22}^{(1)}(\omega)\chi_{31}^{(1),*}(\omega) \\ 2\chi_{11}^{(1)}(\omega)\chi_{33}^{(1),*}(\omega) + 2\chi_{13}^{(1)}(\omega)\chi_{31}^{(1),*}(\omega) & 2\chi_{11}^{(1)}(\omega)\chi_{32}^{(1),*}(\omega) + 2\chi_{12}^{(1)}(\omega)\chi_{31}^{(1),*}(\omega) \\ 2\chi_{11}^{(1)}(\omega)\chi_{23}^{(1),*}(\omega) + 2\chi_{13}^{(1)}(\omega)\chi_{21}^{(1),*}(\omega) & 2\chi_{11}^{(1)}(\omega)\chi_{22}^{(1),*}(\omega) + 2\chi_{12}^{(1)}(\omega)\chi_{21}^{(1),*}(\omega) \end{matrix} \Bigg] \quad (C11)$$

It is noteworthy that $\chi_{ijk}^{(2)}(0,\omega,-\omega)$ in Eq. (C11), which describes the rectification of two THz waves with the frequency $\omega$ and $-\omega$, is conceptually different from the second-order dc nonlinear susceptibility $\chi_{ijk}^{(2),\text{dc}}$. The latter can be obtained by letting $\omega=0$ in Eq. (C9). Like the expression of $\chi_{ijk}^{(2)}(2\omega,\omega,\omega)$ in Eq. (C9), the expression of $\chi_{ijk}^{(2)}(0,\omega,-\omega)$ is only valid when the amplitude of the $E_i^{\text{inc}}$ is not too large. On one hand, the derivation of $\chi_{ijk}^{(2)}(0,\omega,-\omega)$ assumes that the center of polarization oscillation is always at $P_i = P_i^0$ as the premise of the perturbation theory. On the other hand, the resulting d.c. shift component $\Delta \mathbf{P}^{(2)}(0,\omega,-\omega)$ describes a shift of the oscillation center away from the $P_i^0$, which would cause sizable change in the coefficient matrices $\mathbf{K}$ and $\mathbf{C}$ if the magnitude of $\Delta \mathbf{P}^{(2)}(0,\omega,-\omega)$ is sufficiently large.

If the incident electrical field contains two frequencies $\omega_n$ and $\omega_m$ ($\omega_n > \omega_m$), the linear polarization oscillation $\Delta \mathbf{P}^{(1)} = \Delta \mathbf{P}^{(1)}(\omega_n) + \Delta \mathbf{P}^{(1)}(\omega_m)$. Substituting the steady-state different frequencies solution $\Delta \mathbf{P}^{(1)}(\omega_n)$ and $\Delta \mathbf{P}^{(1)}(\omega_m)$ into Eq. (C5b) allows for deriving the steady-state solution of $\Delta \mathbf{P}^{(2)}(\omega_n \pm \omega_m)$ and therefore the $\chi_{ijk}^{(2)}(\omega_n \pm \omega_m, \omega_n, \omega_m)$. To do this, we likewise expand the terms that are contained in the expression $\Delta \mathbf{P}_{\text{II}}^{(1)}$ into the following expression,

$$\Delta P_j^{(1)}(\omega_n)\Delta P_k^{(1)}(\omega_m) = \kappa_0^2 |\chi_{jm}| E_m^{\text{inc},0} \cos(\omega_n t + \varphi^{jm}) |\chi_{kn}| E_n^{\text{inc},0} \cos(\omega_m t + \varphi^{kn})$$

$$= \kappa_0^2 \left(|\chi_{jm}| E_m^{\text{inc},0}\right)\left(|\chi_{kn}| E_n^{\text{inc},0}\right)\left(\frac{1}{2}\cos((\omega_n - \omega_m)t + \varphi^{jm} - \varphi^{kn}) + \frac{1}{2}\cos((\omega_n + \omega_m)t + \varphi^{jm} + \varphi^{kn})\right)$$

$$\equiv \frac{\kappa_0^2}{2} \chi_{jm}^{(1)} \chi_{kn}^{(1),*} E_m^{\text{inc},0} E_n^{\text{inc},0} e^{-i(\omega_n - \omega_m)t} + \frac{\kappa_0^2}{2} \chi_{jm}^{(1)} \chi_{kn}^{(1)} E_m^{\text{inc},0} E_n^{\text{inc},0} e^{-i(\omega_n + \omega_m)t}$$

$$= \Delta \mathbf{P}^{(2)}(\omega_n - \omega_m) + \Delta \mathbf{P}^{(2)}(\omega_n + \omega_m), \tag{C12}$$

Following similar procedures as the derivation of $\chi_{ijk}^{(2)}(2\omega)$ and $\chi_{ijk}^{(2)}(0)$, we can derive that,

$$\chi_{ijk}^{(2)}(\omega_n + \omega_m, \omega_n, \omega_m) = -\kappa_0^2 \sum_{\alpha,\beta,\gamma=1,2,3} C_{\alpha\beta\gamma} \chi_{i\alpha}^{(1)}(\omega_n + \omega_m)[\chi_{\beta j}^{(1)}(\omega_n)\chi_{\gamma k}^{(1)}(\omega_m) + \chi_{\beta j}^{(1)}(\omega_m)\chi_{\gamma k}^{(1)}(\omega_n)] \tag{C13}$$

$$\chi_{ijk}^{(2)}(\omega_n - \omega_m, \omega_n, \omega_m) = -\kappa_0^2 \sum_{\alpha,\beta,\gamma=1,2,3} C_{\alpha\beta\gamma} \chi_{i\alpha}^{(1)}(\omega_n - \omega_m)[\chi_{\beta j}^{(1)}(\omega_n)\chi_{\gamma k}^{(1),*}(\omega_m) + \chi_{\beta j}^{(1),*}(\omega_m)\chi_{\gamma k}^{(1)}(\omega_n)] \tag{C14}$$

**Appendix D Thermodynamic validation and symmetry validation**

*Thermodynamic Validation*

Let us first show that the dc nonlinear susceptibility $\chi_{ijk}^{(2),\text{dc}}$, which can be obtained by letting $\omega=0$ in Eq. (C9), can equivalently be obtained from thermodynamic analyses. To this end, we consider a monodomain ferroelectric material, which has an electric Helmholtz free energy density given as,

$$f(T, P_i, E_i, \varepsilon_{ij}) = g_0(T) + f^{\text{Landau}}(T, P_i) - \frac{1}{2}\kappa_0\kappa_b E_i E_j - E_i P_i + f^{\text{Elast}}(\varepsilon_{ij}, P_i), \tag{D1}$$

where $g_0(T)$ is the Gibbs free energy density of the initial nonequilibrium state with zero spontaneous polarization; the Landau free energy density $f^{\text{Landau}}(T, P_i)$ is a function of temperature and ionic polarization; the elastic energy density $f^{\text{Elast}}(\varepsilon_{ij}, P_i)$ is a function of the total strain $\varepsilon_{ij}$ and ionic polarization, where the total strain $\varepsilon_{ij}$ depends on the mechanical boundary condition of the system. If omitting the depolarization field, the total electric field $E_i$ ($E_j$) in Eq. (D1) is the same as the applied electric field. Minimizing $f(T, P_i, E_i, \varepsilon_{ij})$ with respect to $P_i$ under a constant electric field $E_j$ yields a relationship between $P_i$ and $E_i$, that is,

$$E_i = \left(\frac{\partial(f^{\text{Landau}} + f^{\text{Elast}})}{\partial P_i}\right)_{T, E_i, \varepsilon_{ij}} \tag{D2}$$

Performing Taylor expansion for Eq. (D2), one has,

$$E_i = \sum_{j=1,2,3} \left(\frac{\partial^2 f^{\text{Landau}}}{\partial P_i \partial P_j} + \frac{\partial^2 f^{\text{Elast}}}{\partial P_i \partial P_j}\right) \Delta P_j + \frac{1}{2} \sum_{j,k=1,2,3} \left(\frac{\partial^3 f^{\text{Landau}}}{\partial P_i \partial P_j \partial P_k} + \frac{\partial^3 f^{\text{Elast}}}{\partial P_i \partial P_j \partial P_k}\right) \Delta P_j \Delta P_k + \tag{D3}$$

Eq. (D3) can be rewritten in the matrix form, i.e., $\mathbf{E} = \mathbf{K}\Delta\mathbf{P} + \mathbf{C}\Delta\mathbf{P}_{\text{II}}$ (see Sec. II for the definitions of the coefficient matrices $\mathbf{K}$ and $\mathbf{C}$), or in its expanded form,

$$\begin{bmatrix} E_1 \\ E_2 \\ E_3 \end{bmatrix} = \begin{bmatrix} K_{11} & K_{12} & K_{13} \\ K_{21} & K_{22} & K_{23} \\ K_{31} & K_{32} & K_{33} \end{bmatrix} \begin{bmatrix} \Delta P_1 \\ \Delta P_2 \\ \Delta P_3 \end{bmatrix} + \begin{bmatrix} C_{111} & C_{122} & C_{133} & C_{123} & C_{113} & C_{112} \\ C_{211} & C_{222} & C_{233} & C_{223} & C_{213} & C_{212} \\ C_{311} & C_{322} & C_{333} & C_{323} & C_{313} & C_{312} \end{bmatrix} \begin{bmatrix} \Delta P_1^2 \\ \Delta P_2^2 \\ \Delta P_3^2 \\ 2\Delta P_2 \Delta P_3 \\ 2\Delta P_1 \Delta P_3 \\ 2\Delta P_1 \Delta P_2 \end{bmatrix}, \tag{D4}$$

Since Eqs. (D2-4) are derived for the thermodynamic equilibrium state of polarization under an applied d.c. electric field, the polarization does not change with time. Therefore, the time derivatives of the polarization are zero, i.e., $\partial^n \Delta \mathbf{P}/\partial t^n = 0$, $n=1,2$. Under this condition, Equation (D4) is equivalent to Eq. (3). Accordingly, the $\chi_{ij}^{(1),\text{dc}}$ derived from Eq. (D4), with $\chi^{(1),\text{dc}} = \mathbf{K}^{-1}$ is equivalent to the solution shown in Eq. (4) under $\omega=0$. Likewise, the solution of $\chi_{ijk}^{(2),\text{dc}}$ derived from Eq. (D4) via the perturbation method is equivalent to the solution shown in Eq. (C9) under $\omega=0$, as listed below (where the superscript 'dc' is omitted for brevity).

$$\begin{bmatrix} \chi_{111}^{(2)} & \chi_{122}^{(2)} & \chi_{133}^{(2)} & \chi_{123}^{(2)} & \chi_{113}^{(2)} & \chi_{112}^{(2)} \\ \chi_{211}^{(2)} & \chi_{222}^{(2)} & \chi_{233}^{(2)} & \chi_{223}^{(2)} & \chi_{213}^{(2)} & \chi_{212}^{(2)} \\ \chi_{311}^{(2)} & \chi_{322}^{(2)} & \chi_{333}^{(2)} & \chi_{323}^{(2)} & \chi_{313}^{(2)} & \chi_{312}^{(2)} \end{bmatrix}$$

$$= -\kappa_0^2 \begin{bmatrix} \chi_{11}^{(1)} & \chi_{12}^{(1)} & \chi_{13}^{(1)} \\ \chi_{21}^{(1)} & \chi_{22}^{(1)} & \chi_{23}^{(1)} \\ \chi_{31}^{(1)} & \chi_{32}^{(1)} & \chi_{33}^{(1)} \end{bmatrix} \cdot \begin{bmatrix} C_{111} & C_{122} & C_{133} & C_{123} & C_{113} & C_{112} \\ C_{211} & C_{222} & C_{233} & C_{223} & C_{213} & C_{212} \\ C_{311} & C_{322} & C_{333} & C_{323} & C_{313} & C_{312} \end{bmatrix}.$$

$$\begin{bmatrix}
{\chi_{11}^{(1)}}^2 & {\chi_{12}^{(1)}}^2 & {\chi_{13}^{(1)}}^2 & 2\chi_{12}^{(1)}\chi_{13}^{(1)} & 2\chi_{11}^{(1)}\chi_{13}^{(1)} & 2\chi_{11}^{(1)}\chi_{12}^{(1)} \\
{\chi_{21}^{(1)}}^2 & {\chi_{22}^{(1)}}^2 & {\chi_{23}^{(1)}}^2 & 2\chi_{22}^{(1)}\chi_{23}^{(1)} & 2\chi_{21}^{(1)}\chi_{23}^{(1)} & 2\chi_{21}^{(1)}\chi_{22}^{(1)} \\
{\chi_{31}^{(1)}}^2 & {\chi_{32}^{(1)}}^2 & {\chi_{33}^{(1)}}^2 & 2\chi_{32}^{(1)}\chi_{33}^{(1)} & 2\chi_{31}^{(1)}\chi_{33}^{(1)} & 2\chi_{31}^{(1)}\chi_{32}^{(1)} \\
2\chi_{21}^{(1)}\chi_{31}^{(1)} & 2\chi_{22}^{(1)}\chi_{32}^{(1)} & 2\chi_{23}^{(1)}\chi_{33}^{(1)} & 2\chi_{22}^{(1)}\chi_{33}^{(1)}+2\chi_{23}^{(1)}\chi_{32}^{(1)} & 2\chi_{21}^{(1)}\chi_{33}^{(1)}+2\chi_{23}^{(1)}\chi_{31}^{(1)} & 2\chi_{21}^{(1)}\chi_{32}^{(1)}+2\chi_{22}^{(1)}\chi_{31}^{(1)} \\
2\chi_{11}^{(1)}\chi_{31}^{(1)} & 2\chi_{12}^{(1)}\chi_{32}^{(1)} & 2\chi_{13}^{(1)}\chi_{33}^{(1)} & 2\chi_{12}^{(1)}\chi_{33}^{(1)}+2\chi_{13}^{(1)}\chi_{32}^{(1)} & 2\chi_{11}^{(1)}\chi_{33}^{(1)}+2\chi_{13}^{(1)}\chi_{31}^{(1)} & 2\chi_{11}^{(1)}\chi_{32}^{(1)}+2\chi_{12}^{(1)}\chi_{31}^{(1)} \\
2\chi_{11}^{(1)}\chi_{21}^{(1)} & 2\chi_{12}^{(1)}\chi_{22}^{(1)} & 2\chi_{13}^{(1)}\chi_{23}^{(1)} & 2\chi_{12}^{(1)}\chi_{23}^{(1)}+2\chi_{13}^{(1)}\chi_{22}^{(1)} & 2\chi_{11}^{(1)}\chi_{23}^{(1)}+2\chi_{13}^{(1)}\chi_{21}^{(1)} & 2\chi_{11}^{(1)}\chi_{22}^{(1)}+2\chi_{12}^{(1)}\chi_{21}^{(1)}
\end{bmatrix} \quad (D5)$$

To demonstrate Eq. (D5) is valid, we consider a tetragonal BaTiO$_3$ single crystal with a spontaneous polarization $\mathbf{P}^0 = (0,0,P_3^0)$ under zero electric field ($E_3^{dc}=0$), as shown in Fig. 1(a). The $\chi_{333}^{(2),dc}$ calculated via Eq. (D5) is $-1.573\times10^{-6}$ m/V, while the $\chi_{33}^{(1),dc}$ calculated by setting $\omega=0$ in Eq. (7) is 125.55. In parallel, we evaluate the value of $P_3$ under different bias electric fields $E_3^{dc}$ at thermodynamic equilibrium by numerically minimizing (via the random search method) the corresponding electric Helmholtz free energy density $f(T,P_i,E_i,\varepsilon_{ij})$ in the Mathematica software, and obtain a static $P_3$–$E_3^{dc}$ curve, as shown in Fig. 5. Next, we fit this static curve using the equation $P_3 = P_3^0 + \kappa_0\left(\chi_{33}^{(1),dc}E_3^{dc} + \chi_{333}^{(2),dc}{E_3^{dc}}^2\right)$, through which we determine that $\chi_{33}^{(1),dc}=125.2$ and $\chi_{333}^{(2),dc}=-1.576\times10^{-6}$ m/V, which agree well with the analytically calculated values.

*Symmetry Validation*

The numbers of nonzero and independent elements of both the $\chi_{ij}^{(1),dc}$ and $\chi_{ijk}^{(2),dc}$ can be mathematically evaluated based on Eq. (D5), which depends on the coefficient tensors **C** and **K** (in other words, the symmetry of the $f^{\text{Landau}}$ and $f^{\text{Elast}}$ with respect to the equilibrium polarization state and the mechanical boundary condition of the system). As an example, considering a stress-free boundary condition, the numbers of nonzero(independent) elements of the $\chi_{ij}^{(1),dc}$ for cubic, P4mm, Amm2, and R3m BaTiO$_3$ bulk crystals are determined to be 3(1), 3(2), 5(3), and 9(2), respectively, in the cubic coordinate system. The numbers of nonzero(independent) elements of the $\chi_{ijk}^{(2),dc}$ for cubic, P4mm, Amm2, and R3m BaTiO$_3$ bulk crystals are 0(0), 5(2), 10(3), and 18(3), respectively, in the cubic coordinate system. By transforming the $\chi_{ij}^{(1),dc}$ and the $\chi_{ijk}^{(2),dc}$ tensors from the cubic to the principal coordinate system (PCS), one can diagonalize the $\chi_{ij}^{(1),dc}$ tensor and simplify the matrix for the $\chi_{ijk}^{(2),dc}$ tensor. We perform this transformation by first transforming the polarization $P_j$ in the Landau free energy from the cubic coordinate system (indicated as the $x_1$-$x_2$-$x_3$ coordinates in Table D1) to $P_i'$ in the PCS (indicated as the $x_1'$-$x_2'$-$x_3'$ coordinates in Table D1) via $P_i' = T_{ij}P_j$, and then calculating the $\chi_{ij}'^{(1),dc}$ and the $\chi_{ijk}'^{(2),dc}$ tensors in the PCS based on Eq. (D4) and (D5). The transformation matrix $T_{ij}$ is calculated as,

$$\mathbf{T} = \begin{bmatrix} \cos\varphi_1 & -\sin\varphi_1 & 0 \\ \sin\varphi_1 & \cos\varphi_1 & 0 \\ 0 & 0 & 1 \end{bmatrix} \cdot \begin{bmatrix} 1 & 0 & 0 \\ 0 & \cos\Phi & -\sin\Phi \\ 0 & \sin\Phi & \cos\Phi \end{bmatrix} \cdot \begin{bmatrix} \cos\varphi_2 & -\sin\varphi_2 & 0 \\ \sin\varphi_2 & \cos\varphi_2 & 0 \\ 0 & 0 & 1 \end{bmatrix}, \quad (D6)$$

where the Euler angles $\varphi_1$, $\Phi$, and $\varphi_2$ (also illustrated in Table D1) describe the sequence of rotating the $x_1$-$x_2$-$x_3$ (cubic) to the $x_1'$-$x_2'$-$x_3'$ (PCS). Specifically, the rotation starts with an initial rotation $\varphi_1$ about the $x_3$-axis, followed by a rotation $\Phi$ about the $x_1$-axis, and then a final rotation $\varphi_2$ about the $x_3$-axis again.

As shown in Table D1, the numbers of nonzero(independent) elements of the $\chi'^{(2),dc}_{ijk}$ for cubic, P4mm, Amm2, and R3m BaTiO$_3$ bulk crystals are 0(0), 5(2), 5(3), and 8(3), respectively, in the PCS. The obtained number of independent and non-zero elements for the $\chi'^{(2),dc}_{ijk}$ are consistent with ref. [26].

**Table D1. List of elements in the $\chi_{ij}^{(1),dc}$ and $\chi_{ijk}^{(2),dc}$ of a stress-free BaTiO$_3$ bulk crystal**

| Crystal phase | $\chi_{ij}^{(1)}$ in the cubic and the principal coordinate systems | $\chi_{ijk}^{(2)}$ in the cubic and the principal coordinate systems |
|---|---|---|
| 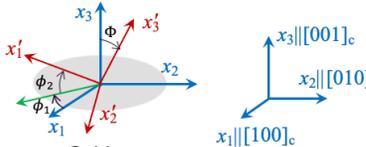 Cubic, $\mathbf{P}^0 = 0$, $(\phi_1, \Phi, \phi_2) = (0°, 0°, 0°)$, T=200 °C, $\mathbf{P}^0 = 0$ | $\begin{pmatrix} \chi_{11}^{(1)} & 0 & 0 \\ 0 & \chi_{11}^{(1)} & 0 \\ 0 & 0 & \chi_{11}^{(1)} \end{pmatrix}$ $\begin{pmatrix} \chi'^{(1)}_{11} & 0 & 0 \\ 0 & \chi'^{(1)}_{11} & 0 \\ 0 & 0 & \chi'^{(1)}_{11} \end{pmatrix}$ $\chi_{11}^{(1)} = \chi'^{(1)}_{11} = 1610.96$ | $\begin{pmatrix} 0 & 0 & 0 & 0 & 0 & 0 \\ 0 & 0 & 0 & 0 & 0 & 0 \\ 0 & 0 & 0 & 0 & 0 & 0 \end{pmatrix}$ $\begin{pmatrix} 0 & 0 & 0 & 0 & 0 & 0 \\ 0 & 0 & 0 & 0 & 0 & 0 \\ 0 & 0 & 0 & 0 & 0 & 0 \end{pmatrix}$ |
| 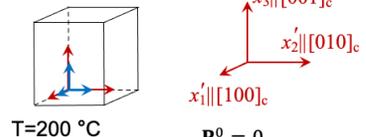 Tetragonal P4mm, $\mathbf{P}^0 = (0,0,P_3^0)$, $(\phi_1, \Phi, \phi_2) = (0°, 0°, 0°)$, T=25 °C, $\mathbf{P}^0 = (0,0,0.26)$ C/m$^3$ | $\begin{pmatrix} \chi_{11}^{(1)} & 0 & 0 \\ 0 & \chi_{11}^{(1)} & 0 \\ 0 & 0 & \chi_{33}^{(1)} \end{pmatrix}$ $\begin{pmatrix} \chi'^{(1)}_{11} & 0 & 0 \\ 0 & \chi'^{(1)}_{11} & 0 \\ 0 & 0 & \chi'^{(1)}_{33} \end{pmatrix}$ $\chi_{11}^{(1)} = \chi'^{(1)}_{11} = 3599.76$ $\chi_{33}^{(1)} = \chi'^{(1)}_{33} = 187.85$ | $\begin{pmatrix} 0 & 0 & 0 & 0 & 2\chi_{311}^{(2)} & 0 \\ 0 & 0 & 0 & 2\chi_{311}^{(2)} & 0 & 0 \\ \chi_{311}^{(2)} & \chi_{311}^{(2)} & \chi_{333}^{(2)} & 0 & 0 & 0 \end{pmatrix}$ $\begin{pmatrix} 0 & 0 & 0 & 0 & 2\chi'^{(2)}_{311} & 0 \\ 0 & 0 & 0 & 2\chi'^{(2)}_{311} & 0 & 0 \\ \chi'^{(2)}_{311} & \chi'^{(2)}_{311} & \chi'^{(2)}_{333} & 0 & 0 & 0 \end{pmatrix}$ $\chi_{311}^{(2)} = \chi'^{(2)}_{311} = 8.74 \times 10^{-5}$ m/V $\chi_{333}^{(2)} = \chi'^{(2)}_{333} = 4.37 \times 10^{-6}$ m/V |
| 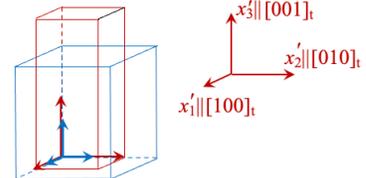 Orthorhombic Amm2, $\mathbf{P}^0 = (P_1^0, 0, P_3^0)$ $P_1^0 = P_3^0$, $(\phi_1, \Phi, \phi_2) = (90°, 45°, -90°)$, T=−10 °C, $\mathbf{P}^0 = (0.225,0,0.225)$ C/m$^3$ | $\begin{pmatrix} \chi_{11}^{(1)} & 0 & \chi_{13}^{(1)} \\ 0 & \chi_{22}^{(1)} & 0 \\ \chi_{13}^{(1)} & 0 & \chi_{11}^{(1)} \end{pmatrix}$ $\begin{pmatrix} \chi'^{(1)}_{11} & 0 & 0 \\ 0 & \chi'^{(1)}_{22} & 0 \\ 0 & 0 & \chi'^{(1)}_{33} \end{pmatrix}$ $\chi_{22}^{(1)} = \chi'^{(1)}_{22} = 2464.09$ $\chi_{11}^{(1)} = 698.7$ $\chi_{13}^{(1)} = -466.1$ $\chi'^{(1)}_{11} = 1164.8$ $\chi'^{(1)}_{33} = 232.6$ | $\begin{pmatrix} \chi_{111}^{(2)} & \chi_{122}^{(2)} & \chi_{133}^{(2)} & 0 & 2\chi_{133}^{(2)} & 0 \\ 0 & 0 & 0 & 2\chi_{133}^{(2)} & 0 & 2\chi_{122}^{(2)} \\ \chi_{122}^{(2)} & \chi_{111}^{(2)} & \chi_{133}^{(2)} & 0 & 2\chi_{122}^{(2)} & 0 \end{pmatrix}$ $\begin{pmatrix} 0 & 0 & 0 & 0 & 2\chi'^{(2)}_{311} & 0 \\ 0 & 0 & 0 & 2\chi'^{(2)}_{322} & 0 & 0 \\ \chi'^{(2)}_{311} & \chi'^{(2)}_{322} & \chi'^{(2)}_{333} & 0 & 0 & 0 \end{pmatrix}$ $\chi_{111}^{(2)} = 7.65 \times 10^{-5}$ m/V $\chi_{122}^{(2)} = 3.85 \times 10^{-5}$ m/V $\chi_{133}^{(2)} = -2.32 \times 10^{-5}$ m/V $\chi'^{(2)}_{311} = 7.05 \times 10^{-5}$ m/V $\chi'^{(2)}_{322} = 5.45 \times 10^{-5}$ m/V $\chi'^{(2)}_{333} = 4.92 \times 10^{-6}$ m/V |
| 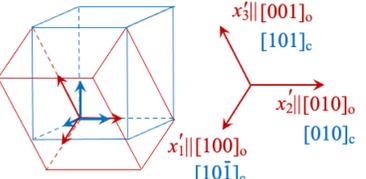 Rhombohedral R3m, $\mathbf{P}^0 = (P_1^0, P_2^0, P_3^0)$ $P_1^0 = P_2^0 = P_3^0$, $(\phi_1, \Phi, \phi_2) = (0°, 54.736°, 45°)$, T=−100 °C, $\mathbf{P}^0 = (0.221,0.221,0.221)$ C/m$^3$ | $\begin{pmatrix} \chi_{11}^{(1)} & \chi_{12}^{(1)} & \chi_{12}^{(1)} \\ \chi_{12}^{(1)} & \chi_{11}^{(1)} & \chi_{12}^{(1)} \\ \chi_{12}^{(1)} & \chi_{12}^{(1)} & \chi_{11}^{(1)} \end{pmatrix}$ $\begin{pmatrix} \chi'^{(1)}_{11} & 0 & 0 \\ 0 & \chi'^{(1)}_{11} & 0 \\ 0 & 0 & \chi'^{(1)}_{33} \end{pmatrix}$ $\chi_{11}^{(1)} = 636.58$ $\chi_{12}^{(1)} = -237.93$ $\chi'^{(1)}_{11} = 874.51$ $\chi'^{(1)}_{33} = 160.72$ | $\begin{pmatrix} \chi_{111}^{(2)} & \chi_{122}^{(2)} & \chi_{122}^{(2)} & \chi_{123}^{(2)} & 2\chi_{122}^{(2)} & 2\chi_{122}^{(2)} \\ \chi_{122}^{(2)} & \chi_{111}^{(2)} & \chi_{122}^{(2)} & 2\chi_{122}^{(2)} & \chi_{123}^{(2)} & 2\chi_{122}^{(2)} \\ \chi_{122}^{(2)} & \chi_{122}^{(2)} & \chi_{111}^{(2)} & 2\chi_{122}^{(2)} & 2\chi_{122}^{(2)} & \chi_{123}^{(2)} \end{pmatrix}$ $\begin{pmatrix} 0 & 0 & 0 & 0 & 2\chi'^{(2)}_{311} & -2\chi'^{(2)}_{222} \\ -\chi'^{(2)}_{222} & \chi'^{(2)}_{222} & 0 & 2\chi'^{(2)}_{311} & 0 & 0 \\ \chi'^{(2)}_{311} & \chi'^{(2)}_{311} & \chi'^{(2)}_{333} & 0 & 0 & 0 \end{pmatrix}$ $\chi_{111}^{(2)} = 6.66 \times 10^{-5}$ m/V $\chi_{122}^{(2)} = -1.65 \times 10^{-5}$ m/V $\chi_{123}^{(2)} = 3.58 \times 10^{-5}$ m/V $\chi'^{(2)}_{222} = -6.21 \times 10^{-5}$ m/V $\chi'^{(2)}_{311} = 2.81 \times 10^{-5}$ m/V $\chi'^{(2)}_{333} = 1.87 \times 10^{-6}$ m/V |

**Appendix E. Evaluating the $\left|\chi_{333}^{(2)}\right|(2\omega)$ and $\chi_{33}^{(1),\text{Im}}(\omega)$ from dynamical phase-field simulations**

A dynamical phase-field model with coupled dynamics of strain, polarization, and EM waves [18,25] was used to simulate the excitation of polarization oscillation in a freestanding tetragonal BaTiO$_3$ slab by a monochromatic continuous THz wave with an angular frequency $\omega$. The BaTiO$_3$ slab has an initial equilibrium polarization $(P_1^0, P_2^0, P_3^0)$=(0, 0, 0.26 C/m$^2$) at 298 K. The slab thickness is set to be 10 nm (which is far smaller than the THz wavelength) to ensure that the excited polarization is spatially uniform along the thickness direction and in-phase. A sinusoidal source current $J_x = J_x^0 \sin(\omega t)$ was injected to generate a continuous incident THz wave with an electric field component that is spatially uniform in the thin BaTiO$_3$ slab, i.e., $E_x(t) = E_x^0 e^{-i\omega t}$, with $x \equiv 3$ (see Fig. 1(a)). We consider 36 different $\omega$ in total near the two peaks of the analytically calculated $\left|\chi_{333}^{(2)}\right|(2\omega)$ shown in Fig. 2(a) and performed 36 groups of simulations to extract the $\left|\chi_{333}^{(2)}\right|(2\omega)$ at each $\omega$. In all simulations, we set $J_x^0$=2×10$^{12}$ A/m$^2$, leading to an $E_x^{\text{inc},0}$ of 376730 V/m. Under this electric field, the BaTiO$_3$ slab will be driven into the anharmonic regime, but the magnitude of the dc polarization shift $\Delta P_3^{(2)}(0)$ arising from $\chi_{333}^{(2)}(0,\omega,-\omega)$, see Eq. (C11), remains relatively small. For such a weakly nonlinear oscillation, one can expect that the numerical simulation results would agree well with the results calculated analytically based on the perturbation theory, because a negligibly small $\Delta P_i^{(2)}(0)$ is the premise of the present analytical model, as discussed in Appendix C (see the paragraph before Eq. (C4)). Details of the dynamical phase-field model and the set-up of the numerical simulations are provided in [25].

As an example, Figure 6(a) shows the steady-state evolution of the $\Delta P_3(t)$ in the middle layer of the BaTiO$_3$ slab (note that $\Delta P_3$ is spatially uniform in the slab) under a continuous THz wave excitation with $\omega$=0.5$\omega_3$=2$\pi$×2.0542 THz, corresponding to the first peak of $\left|\chi_{333}^{(2)}\right|(2\omega)$ in Fig. 1(a). Figure 6(b) shows the frequency spectrum of $\Delta P_3(t)$, which display two prominent peaks. The first peak at $\omega/2\pi$=2.0542 THz is related to the linear polarization oscillation $\Delta P_3^{(1)}(t) = \Delta P_3^{(1),0} e^{i(-\omega t + \varphi^{(1)})}$ that has the same frequency as the incident THz wave. The second peak at $\omega/2\pi$=4.1084 THz corresponds to the second-harmonic component of the polarization oscillation, given by $\Delta P_3^{(2)}(t) = \Delta P_3^{(2),0} e^{i(-2\omega t + \varphi^{(2)})}$. Performing the inverse Fourier transform for the second peak allows for reconstructing the temporal profile of the $\Delta P_3^{(2)}(t)$, as shown in Fig. 6(c), from which both the amplitude $\Delta P_3^{(2),0}$ (=0.027762 mC/m$^2$) and phase $\varphi^{(2)}$ (= 1.48140 rad) (84.878°) can be extracted. Thus, $\left|\chi_{333}^{(2)}(2\omega)\right| = 2\Delta P_3^{(2),0}/\kappa_0 \left(E_3^{\text{inc},0}\right)^2$ is calculated to be 4.454×10$^{-5}$ m/V, which agrees well with the analytically calculated value of 4.67×10$^{-5}$ m/V. Likewise, performing the inverse Fourier transform for the first peak enables the reconstruction of $\Delta P_x^{(1)}(t)$, from which we obtain $\Delta P_3^{(1),0}$=0.569 mC/m$^2$ and $\varphi^{(1)}$ = 0.0515 rad (2.951°). Therefore, $\left|\chi_{33}^{(1)}(\omega)\right|$ = $\Delta P_3^{(1),0}/\kappa_0 E_3^{\text{inc},0}$ = 170.5995. Given that $\chi_{33}^{(1)} = \left|\chi_{33}^{(1)}\right| e^{i\varphi^{(1)}} = \chi_{33}^{(1),\text{Re}} + i\chi_{33}^{(1),\text{Im}}$, one has $\chi_{33}^{(1),\text{Im}} = \left|\chi_{33}^{(1)}\right| \sin\varphi^{(1)}$ =8.651, which agrees well with the analytically calculated $\chi_{33}^{(1),\text{Im}}$ of 7.259 at $\omega$=2$\pi$×2.0542 THz.

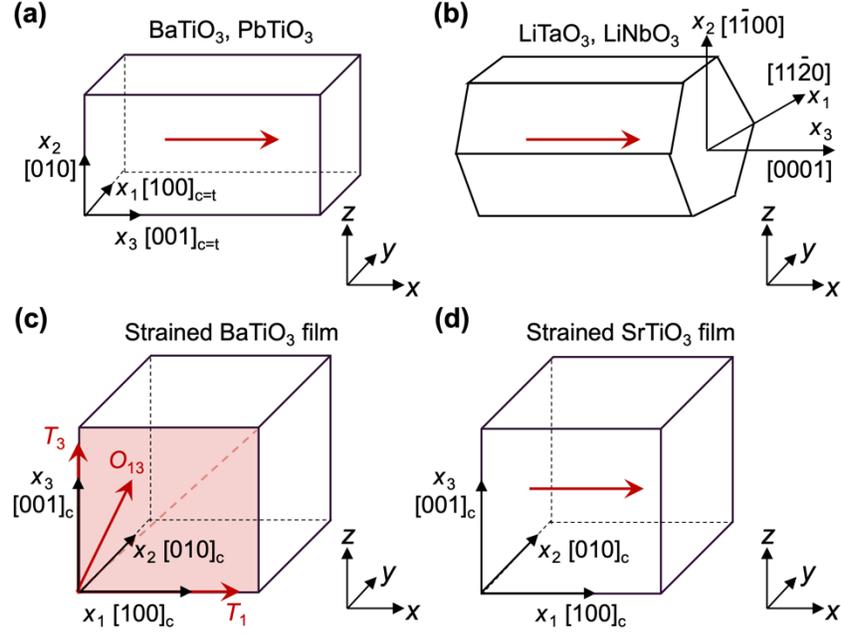

**Figure 1**. Schematics of (**a**) the tetragonal unit cells of BaTiO$_3$ and PbTiO$_3$; (**b**) the hexagonal unit cells of the LiTaO$_3$ and LiNbO$_3$. (**c,d**) Cubic representation of the pseudocubic (pc) unit cells of a (001)$_{pc}$ strained BaTiO$_3$ film and a (001)$_{pc}$ strained SrTiO$_3$ film. The red arrow indicates the direction of the spontaneous polarization. In (**c,d**), $x_1$-$x_2$-$x_3$ coordinates indicate the Cartesian axes of the cubic paraelectric unit cells of the BaTiO$_3$ and SrTiO$_3$, consistent with the notation $P_i$ ($i$=1,2,3) in the LGD thermodynamic energy density function. $x$-$y$-$z$ coordinates indicate the lab coordinate system.

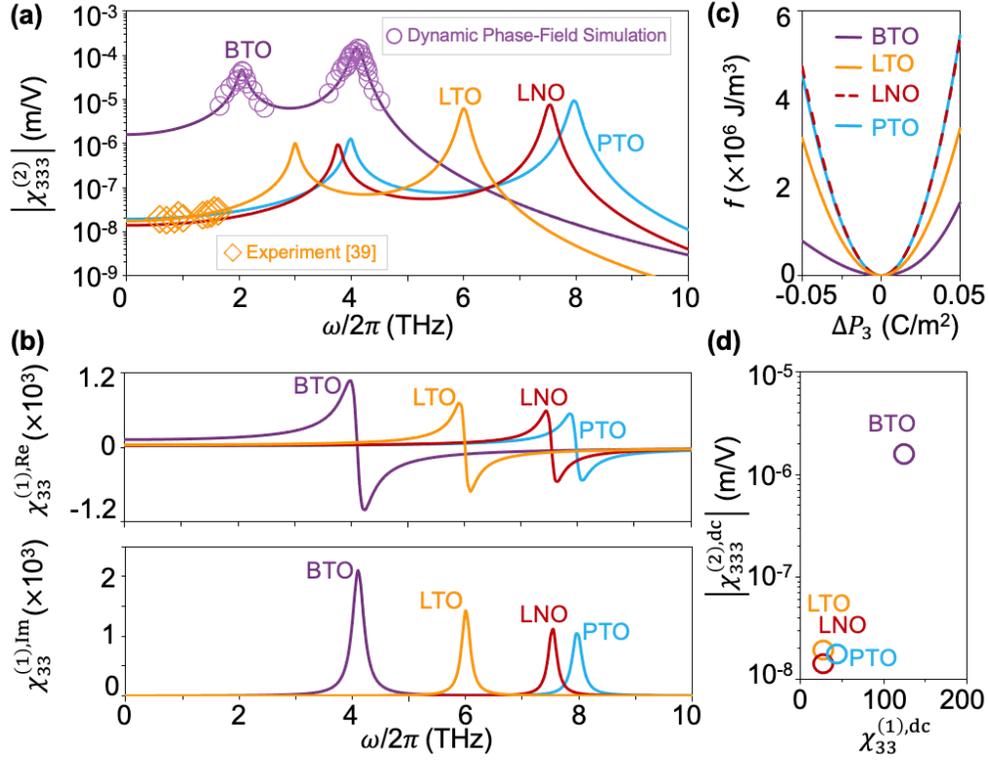

**Figure 2**. Frequency-dependent (**a**) modulus of the nonlinear susceptibility $\left|\chi^{(2)}_{333}\right|$; (**b**) Real and imaginary component of the linear susceptibility $\chi^{(1)}_{33}$ in tetragonal BaTiO$_3$ (BTO) and PbTiO$_3$ (PTO) as well as the trigonal LiTaO$_3$ (LTO) and LiNbO$_3$ (LNO) bulk ferroelectric crystals. (**c**) Free energy density $f$ as a function of $\Delta P_3 = P_3 - P_3^0$ in these four materials. (**d**) dc linear susceptibility $\chi^{(1),dc}_{33}$ and the magnitude of the dc second-order nonlinear susceptibility $\left|\chi^{(2),dc}_{333}\right|$ of these four materials. The temperature is 298 K, which is below the Curie temperature of the four ferroelectric materials studied herein.

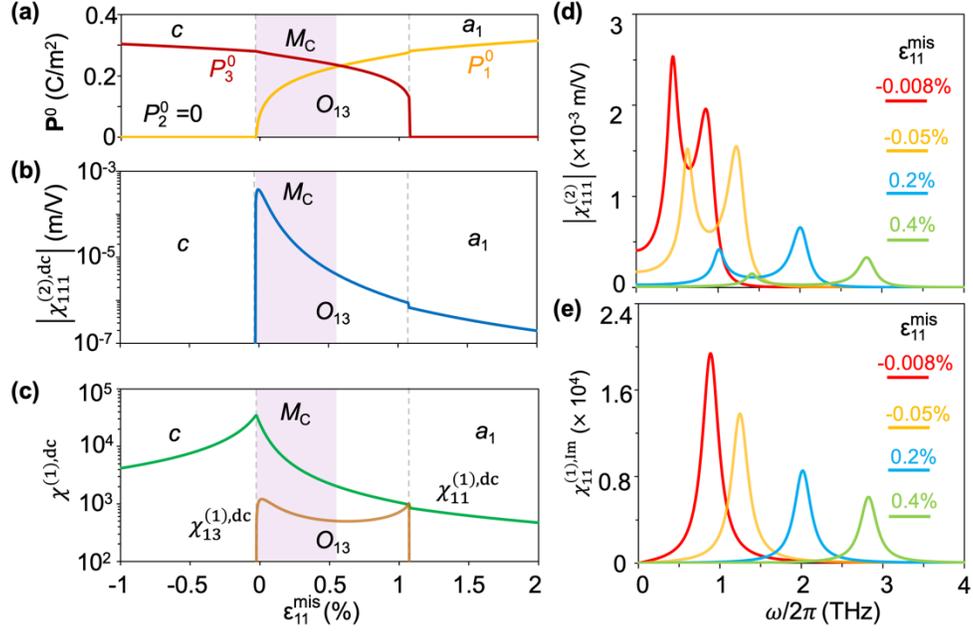

**Figure 3**. Strain-dependent (**a**) equilibrium polarization state $\mathbf{P}^0$; (**b**) dc nonlinear susceptibility absolute value $\left|\chi_{111}^{(2),dc}\right|$; and (**c**) dc nonlinear susceptibility $\chi_{11}^{(1),dc}$ and $\chi_{13}^{(1),dc}$ in a coherently strained $(001)_{pc}$ BaTiO$_3$ film. $\varepsilon_{22}^{mis}$ is fixed at 1%. The shade indicates the monoclinic $M_C$ phase, which belongs to the $O_{13}$ phase but with $\left|P_3^0\right| > \left|P_1^0\right|$. Frequency-dependent (**d**) $\left|\chi_{111}^{(2)}\right|$ and (**e**) $\chi_{11}^{(1),Im}$ under different mismatch strains $\varepsilon_{11}^{mis}$. The temperature is 298 K.

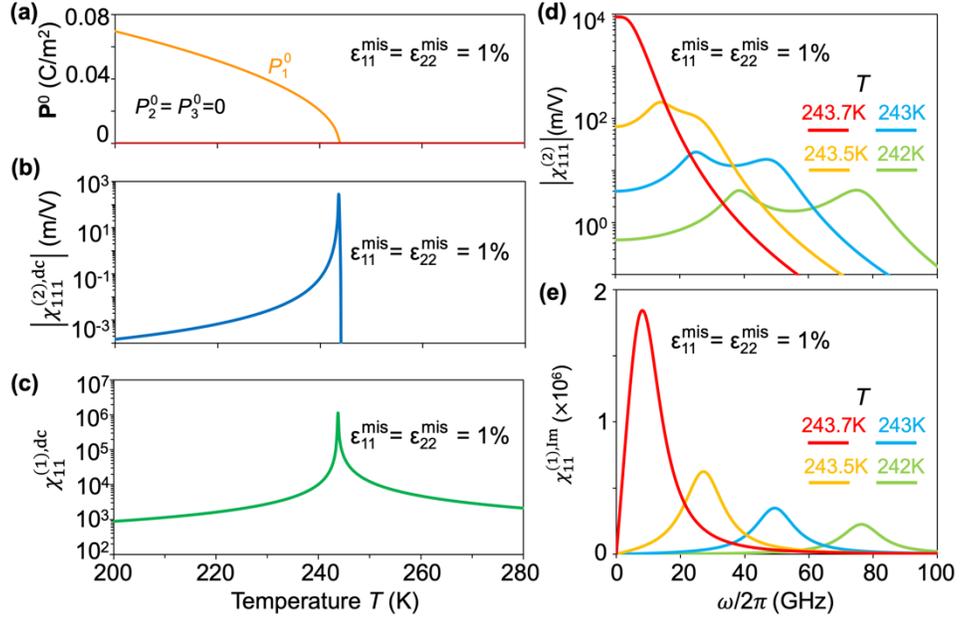

**Figure 4**. Temperature-dependent (**a**) equilibrium polarization state $\mathbf{P}^0$, (**b**) dc nonlinear susceptibility absolute value $\left|\chi_{111}^{(2),dc}\right|$, and (**c**) dc linear susceptibility $\chi_{11}^{(1),dc}$ in a coherently strained $(001)_{pc}$ SrTiO$_3$ film at $\varepsilon_{11}^{mis} = \varepsilon_{22}^{mis}$ =1%. Frequency-dependent (**d**) $\left|\chi_{111}^{(2)}\right|$ and (**e**) $\chi_{11}^{(1),Im}$ under different temperature at $\varepsilon_{11}^{mis}=\varepsilon_{22}^{mis}$=1%. The structural order parameter $q_1=q_2=q_3=0$ under these strain and temperature conditions.

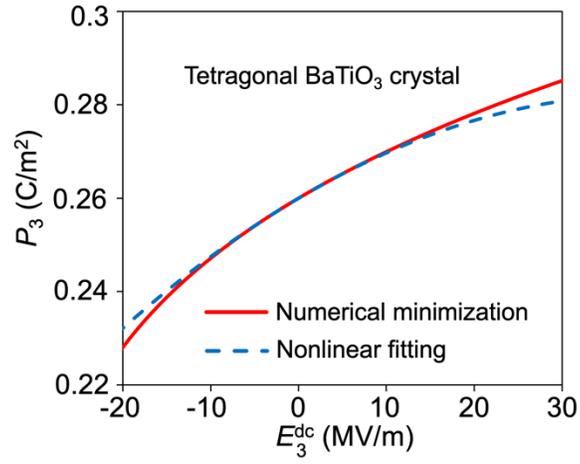

**Figure 5**. The $P_3$ as a function of an applied dc electric field $E_3^{dc}$ obtained by thermodynamic analysis in a bulk tetragonal BaTiO$_3$ single crystal (shown in Fig. 1(a)) at 25°C and its nonlinear fitting as explained in Appendix C.

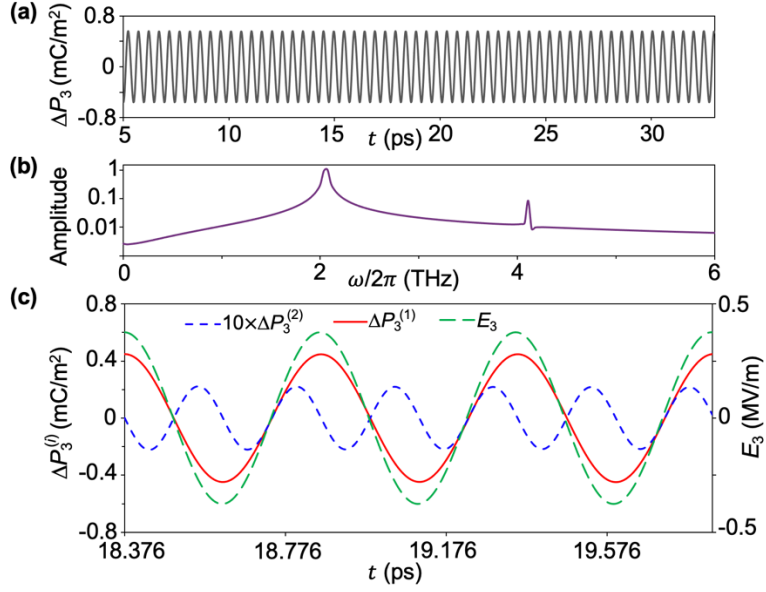

**Figure 6. (a)** Steady-state evolution of the dynamically excited polarization $\Delta P_3(t) = P_3(t) - P_3^0$ in a thin freestanding (100) BaTiO$_3$ slab under a continuous THz wave with $\omega=2\pi\times2.0542$ THz at 298 K; **(b)** Frequency spectrum of the $\Delta P_3(t)$ for the duration of $t$=5-33 ps; **(c)** Reconstructed temporal profiles of $\Delta P_3^{(1)}$ and $\Delta P_3^{(2)}$, as well as the profile of the incident THz electric field $E_3(t)$ obtained from a reference simulation without the (100) BaTiO$_3$ slab. $t$=0 ps is the moment the source current is injected.